# Experimental and Theoretical Study of Flow Dynamics of Expanding Hydrogen Plasma from an ECR Plasma Source

[1]Priti Singh, [1,3] Ramesh Narayanan, [1]Ashish Ganguli, [1]Debaprasad Sahu, [2] Subhasish Bag
[1] Department of Energy Science and Engineering, IIT Delhi, New Delhi, India
[2] Asia Pacific Center for Theoretical Physics, Pohang, Gyeongbuk, Republic of Korea
[3] Corresponding author: rams@dese.iitd.ac.in

## Abstract

The flow of moderately dense ($n \approx 10^{11}$ cm$^{-3}$) hydrogen plasma with high electron temperature ($T_e \approx 30 - 35$ eV) and plasma potential ($V_p \approx 100$ V), expanding towards an expansion chamber (EC) from an ECR plasma source is investigated using Langmuir probes (LPs) and Ion Energy Analysers (IEAs). The LPs reveal presence of a *second ionization zone* (SIZ) a few centimetres from the ECR zone, indicating very short ionization mean paths, which are confirmed theoretically. The SIZ gives a *density boost*, produces a *double layer*, DL ($\Delta V_{DL} \approx 106 - 40$ V) and a *warm electron population after the DL* ($\frac{n_w}{n} \approx 0.04 - 0.10$, $T_w \approx 40 - 60$ eV). IEA measurements conducted at two sites ($z_1 \approx 5$ cm, $z_2 \approx 10$ cm) reveal ions with energies in the range $\approx 65 - 8$ eV at pressures $\approx 1 - 8$ mTorr. DL formation is explained in terms of the SIZ and warm electron generation is analysed in detail.

## 1. Introduction

Hydrogen plasmas play an important role both for industry and research, which emphasises the need for efficient hydrogen plasma sources with capability for producing high density uniform plasmas, with capacity to scale over a wide range of volumes and pressures. It is also well known that great significance of hydrogen (H) plasmas is due to their chemical reactivity and versatility in industrial and research contexts. One prominent application lies in the removal of impurities from metal and semiconductor surfaces, where reactive hydrogen species interact with surface contaminants. These interactions typically result in the formation of volatile compounds such as water ($H_2O$) or hydrocarbons (e.g., $CH_4$), which can then be effectively removed from the system, thereby achieving surface purification.[1-9] In the semiconductor industry, hydrogen plasmas are also extensively utilized for the passivation of semiconductor surfaces. Such passivation significantly enhances the electrical performance and stability of semiconductor devices, making it a critical step in the fabrication of high-efficiency electronic and optoelectronic components.[10-11] It has been demonstrated that diamond films grown in hydrogen-diluted plasma exhibit superior crystallinity, with growth rates in the range of 200–250 μm/h - nearly an order of magnitude higher than conventional CVD techniques. This remarkable enhancement in both growth rate and film quality underscores the critical role of atomic hydrogen in facilitating the formation of high-quality diamond films in such plasma environments.[11] Hydrogen plasma consists of different ion species ($H^+$, $H_2^+$, $H_3^+$, $H^-$, *etc.*), which has led to hydrogen becoming one of the most preferred utility gases for diverse industrial applications.[12-15] In fact, proton ($H^+$) beams have major applications in the medical field as charged-particle radiotherapy.[16,17] Both $H^+$ and $H^-$ ions can also be used for the ion projection lithography.[18]

During the past several years, $H^-$ ion sources have also found a wide range of applications such as in high-energy accelerators,[19] surface modification of materials by ion implantation,[20] semiconductor fabrication,[21] and neutral-beam heating of fusion plasmas.[22,23]

Mostly ECR and RF sources are used to produce hydrogen plasmas. RF sources work effectively at comparatively higher pressures (for enhancing collisions) whereas ECR plasma sources yield fairly high plasma densities at lower pressures; they also help save on the consumption of the operating gases, which is a major factor for industry.

Despite their advantages ECR plasma sources do not find widespread acceptance in industry. Expensive microwave hardware, limitations in scaling to large volumes and inhomogeneities in plasma parameters arising from magnetic fields could be some factors that limit use of ECR based sources in industry. To overcome some of these drawbacks the Plasma Lab at IIT Delhi developed the Compact ECR Plasma Source (CEPS),[24-30] that weighs about ≈ 14 kgs (including the NdFeB permanent magnets) and has length ≈ 60 cm, with the magnetron, triple stub tuner, waveguide adapter, plasma source section, etc., all integrated into a single unit. The microwave hardware is fabricated from aluminium, which is both lightweight and low cost. The magnetrons (2.45 GHz, 800 W, CW) for these sources are those used in commercial microwave ovens and hence are relatively inexpensive. To conclude, the whole principle behind the development of the CEPS was to develop an ECR plasma source that is cheap, compact, portable, versatile, and easy to use. It can be attached to any port of a plasma vessel to fill it with plasma. In addition, the unique magnetic field topology of the NdFeB ring magnets affords a higher electron confinement time (within the ECR zone) and hence, more efficient electron heating and plasma ejection. Proof of the latter was seen in a recent study, where the CEPS was mounted coaxially onto a moderately large volume plasma system (the MVPS).[31] In this study, high energy (~ 95 eV) ion beams were detected in front of the source mouth in the MVPS expansion chamber (EC). It is believed that such energetic ions could only be formed by special nonlinear structures like double layers (DLs), etc. However, a detailed measurement of plasma parameters within the source region was not possible because of rapid burnouts of the Langmuir probes (LPs) due to bombardment by the high energy ions. To unravel the details of plasma generation and ion acceleration within the plasma source section (PSS) therefore, an alternative plasma source was needed that would reproduce the key features of the CEPS experiment, but without the latter's disadvantage of producing severely harsh plasma environments within the PSS. This would enable the plasma to be probed by suitable diagnostics. Such an alternative plasma source, called the *Generic ECR Plasma Source* (GEPS), was developed for the present experiments. For purposes of comparison, the GEPS was mounted on the same expansion chamber (the MVPS) as the CEPS. It may be mentioned that an important aim of the present work was to understand the flow dynamics of the plasma as it escapes from within the PSS into the EC as well as seek a comparison of the CEPS and the GEPS with regard to their performance as plasma sources.

The magnetic field of the GEPS is produced by a *combination of ring magnets and an electromagnet coil* mounted coaxially on the plasma source section of the GEPS. For ease of reference, the magnetic field topology of the CEPS is labeled MF1, while that of the GEPS is labeled MF2.

Analysis, based on measurements to be presented later, reveals the following picture. As electrons, heated by the resonant microwaves, escape the ECR zone they leave behind a plasma with high plasma potential in the source region. The latter arrests the escape of the electrons while driving out the ions. In steady state, both electrons and ions leave together by ambipolar flow towards the mouth of the plasma source section (PSS). Along the way, the hot electrons initiate a second ionization zone (SIZ) that also aids formation of a DL. The former helps boost the plasma density (at the cost of a fall in $T_e$) while the latter accelerates the ions to high energies. The possibility of an SIZ was confirmed theoretically. Concomitantly, a warm electron population is also born that is easily detected downstream in the expansion chamber. Ion energies, measured using a Retarding Field Energy Analyser (RFEA), reveal fairly high ion energies accelerated by the DL. This and other important findings from the experiments are supported using suitable theoretical models.

The paper is organised as follows. A detailed description of the GEPS and its magnetic field structure together with the experimental setup is presented in section 2. Also described in section 2 are the plasma diagnostics used for the experiments. Section 3 presents a detailed discussion of the experimental results along with relevant theoretical models to support the experiments, including a discussion on the formation of the warm electron population. Section 4 presents a brief comparison of the CEPS and GEPS plasmas. Section 5 is the concluding section. An Appendix is included to augment some the theoretical calculations.

## 2 Experimental Setup

### 2.1 The GEPS and its Magnetic Field

The schematic of the basic GEPS system is shown in Fig. 1. It comprises a plasma source section, PSS (ID $\approx$ 9.1 cm) mounted coaxially onto an expansion chamber, EC (ID $\approx$ 50 cm, Length $\approx$ 75 cm); EC is also labeled the Medium Volume Plasma System (MVPS) to distinguish it from the Small and Large Volume Plasma Systems (SVPS, LVPS) in which numerous studies using single and multiple CEPS have been conducted.[25,27] It may be mentioned that the CEPS, in combination with the MVPS, has also been used for argon and hydrogen related studies.[28,30,31] As noted above, though the CEPS and GEPS have the same PSS and EC (the MVPS), the magnet assembly for the two are very different. The CEPS uses a set of three, compact NdFeB ring magnets; detailed discussions of the field topology along with its role in electron confinement and heating can be found in Ref [31].

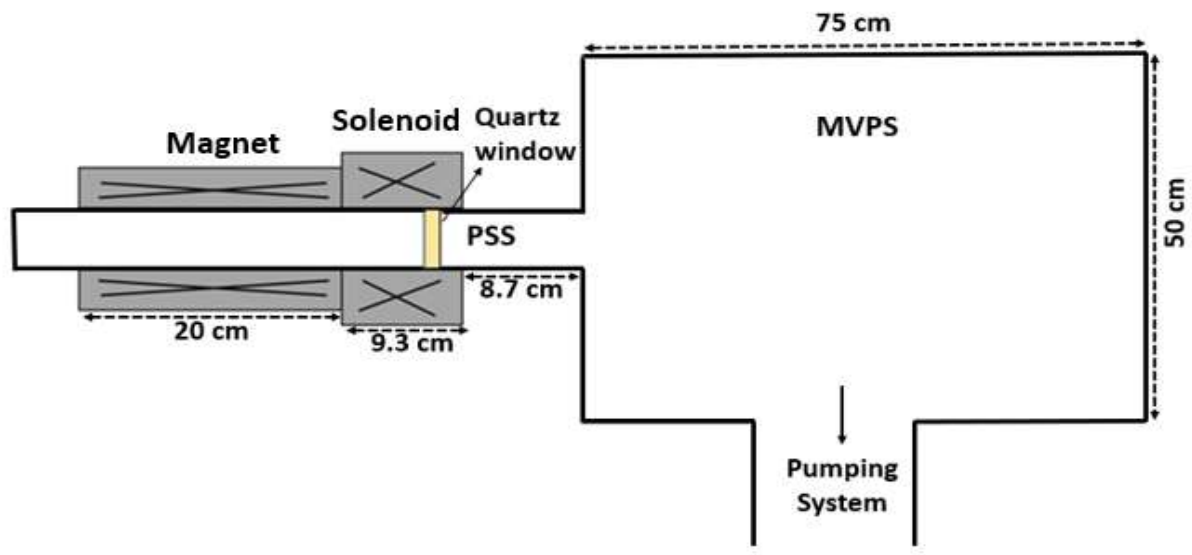

**FIG. 1. Schematic of details of GEPS and its mounting on to the MVPS**

Fig. 1 also shows the magnet assembly for the GEPS, which is a combination of 5 NdFeB ring magnets (ID = 11.6 cm and OD = 17.4 cm) and a solenoid coil (coil ID = 127.2 mm, coil OD = 257.2 mm, wire diameter = 2.6 mm, number of turns: 575), all mounted coaxially on the PSS of the GEPS. Microwave coupling is via the quartz window (Fig. 1). To summarise, the GEPS consists of three basic components: (i) magnetron and microwave coupling; (ii) magnet-coil assembly and (iii) ECR plasma source section (PSS).

Fig. 2(a) gives a half-sectional view of the PSS and the EC (from the axis to the sidewall) along with placement of the magnets and the coil. Also shown are the constant ***B*** contours (solid-coloured lines) within the PSS and the EC along with plots of the magnetic field lines (dashed lines) obtained using COMSOL Multi-physics software and verified experimentally using a three-axis Hall probe [LakeShore, Model 4060]. The ECR contour (or surface in 3D; ≈ 876 G) is shown in red. It can be seen that there are three ECR zones: two of these are at atmospheric pressure (in air) *to the left of the quartz window* that protects the vacuum to its right, and one *after the window in vacuum*, close to the mouth (exit point) of the PSS. The latter arrangement precludes spurious plasma formation by the first two ECR zones, far away from the PSS exit. The radial profile of the ECR contour shows a very weak curvature, implying that in 3D it would resemble a very slightly curved disc. This contrasts greatly with the ECR contour of the CEPS magnets, which forms a bowl in 3D (with its axis along the system axis).[31] The latter generates a curved magnetic mirror profile, providing additional confinement to the electrons. The GEPS field however, offers no such confinement.

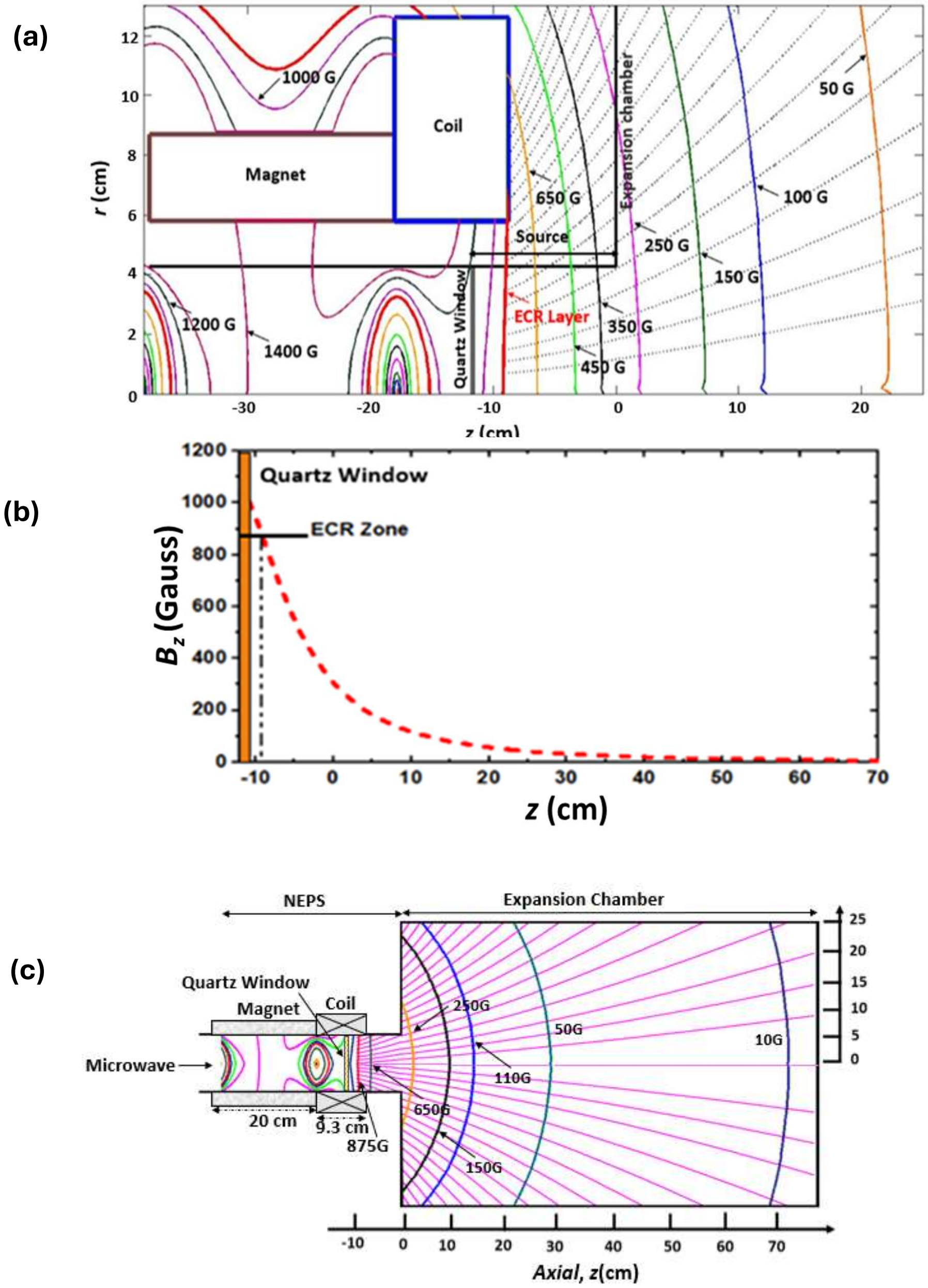

**FIG. 2. Detailed view of the magnetic field topology of the GEPS. (a) The magnetic field structure inside the source section. The solid, coloured lines indicate constant magnetic field contours and the dashed lines, the magnetic field lines; (b) Monotonically decreasing axial magnetic field profile; (c) Magnetic field lines and constant magnetic field contours inside the plasma source section (PSS) and the expansion chamber (MVPS).**

The on-axis variation of $B_z$ in the GEPS is shown in Fig. 2(b). It is seen clearly that $B_z$ decreases monotonically all the way to the chamber end. $B_z \approx 1100$ G at the quartz window; the ECR zone is close to the quartz window (at $z \approx -9.1$ cm) and the on-axis magnetic null that is present for the CEPS is absent. More details of the experimental system can be found in Ref [31].

A view of the field lines and the constant field contours within the PSS and the EC is shown in Fig. 2(c). The field is a simple diverging field that offers no additional confinement to the plasma electrons. A 3D view of the field lines is shown in Fig. 3. For ease of visualization, the ECR zone (≈ 876 G) is shown as an *extended plane.* The field lines intersect it at different ($r$, $z$). On the axis, the field line intersects the ECR plane at $z \approx -9.1$ cm. Away from the axis, the lines exhibit a slight bending towards increasing $r$ as they continue along $z$, indicating diverging field lines. The field decreases monotonically and unlike that for the CEPS it does not pass through a minimum to intersect the ECR plane again to allow particles to revisit the resonant field.

**2.2 Langmuir Probes**

Axial and radial Langmuir probes (LPs) were used to determine the plasma parameters, *viz*., bulk plasma density ($n_e$), bulk electron temperature ($T_e$), the warm electron density and temperature when present ($n_w$, $T_w$) and the plasma potential ($V_p$).[29,32,33] Although, hydrogen is a molecular gas and can form different ions ($H^+, H_2^+, H_3^+$, etc.), it will be assumed that only $H^+$ ions are present here.[28] While the radial LPs had cylindrical probe tips (dia ≈ 0.25 mm, length ≈ 4 mm), the axial probe had a special structure to protect its tip from the bombarding ions. Fig. 4 shows the probe tip to be covered by a protective, ceramic cap and particle collection taking place from the exposed tungsten part, *across* the magnetic field.

The rest of the probe is again covered and insulated by ceramic capillaries. It is worth mentioning here that, even with such a probe design, it was not possible to probe the PSS in the CEPS hydrogen plasma experiments.[31]

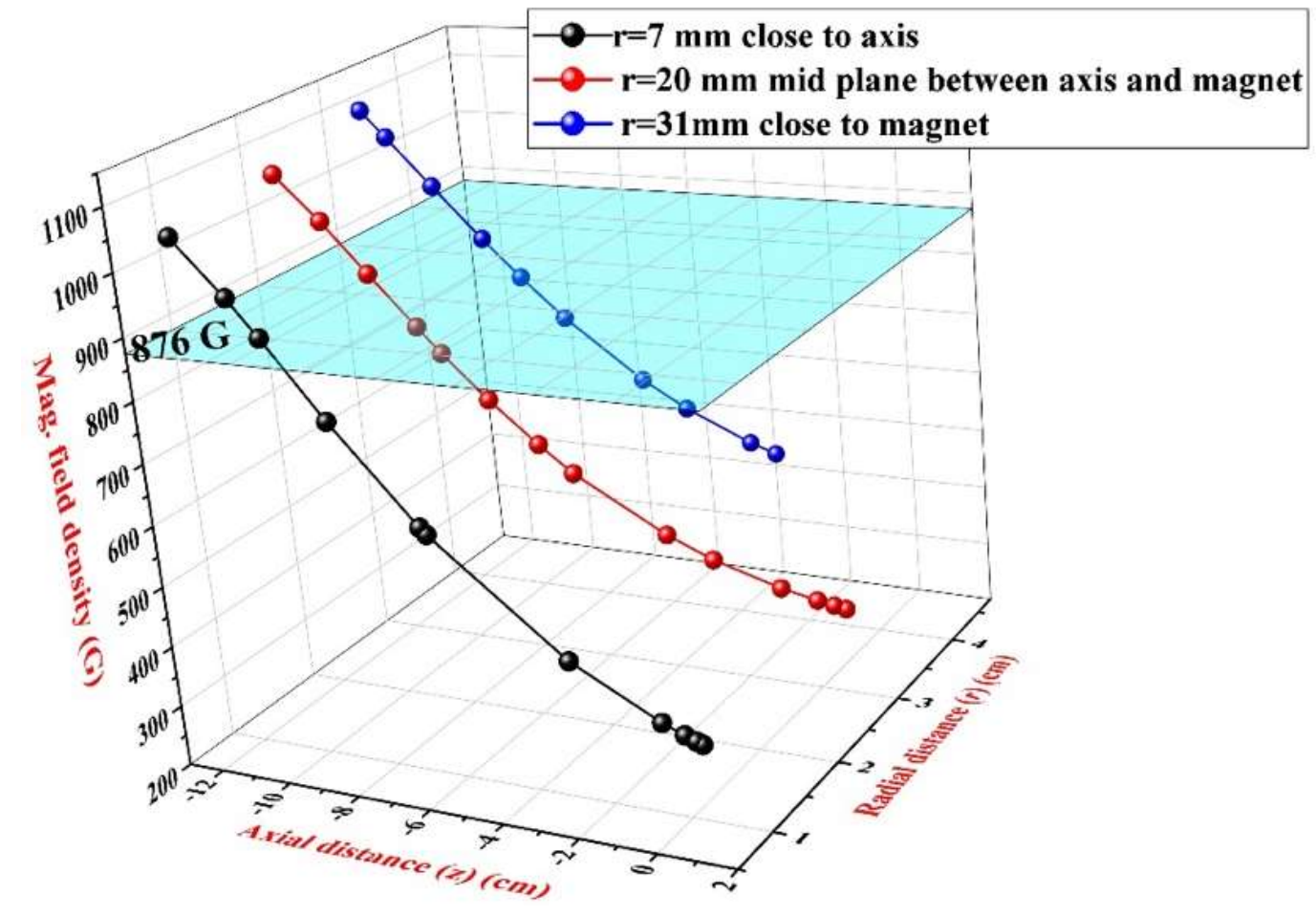

**FIG. 3. 3D magnetic field topology of the GEPS**

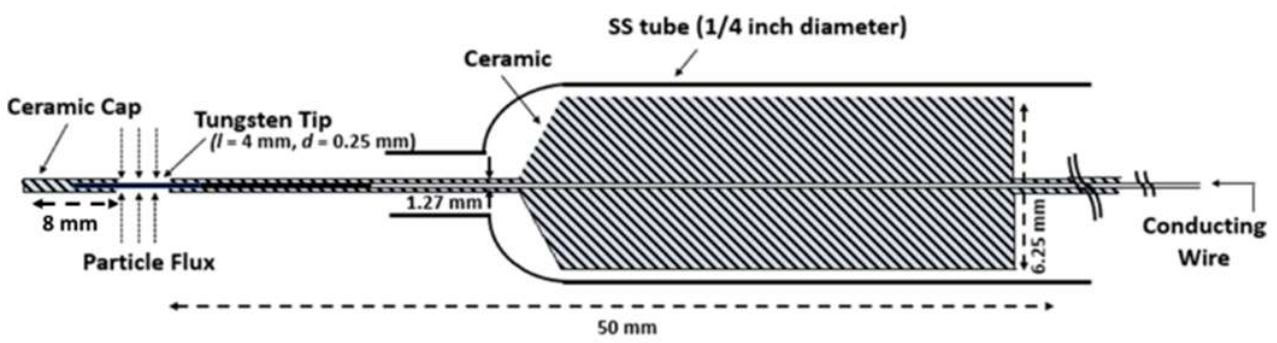

**FIG. 4. Schematic of Axial Langmuir probe[31]**

The *I-V* data was acquired by biasing the probe with a voltage sweep ranging from – 180 V to + 50 V in 500 steps within a time span of ≈ 500 msec. The resulting probe current measured as a function of the probe bias voltage was recorded using the analog inputs of a National data acquisition card, NI 6143 S series (16-Bit, 250 kS/s/ch). The details of the LP data collection methodology are given in the authors' recent paper.[31]

***Typical LP Characteristics:*** In the present work in addition to the usual single temperature electron populations, two electron populations are also found fairly often. Typically, in such cases one obtains a low temperature, high-density bulk population along with a low density, high temperature, warm population.

***Single Electron Temperature Population:*** Fig. 5(a) gives typical $I - V$ characteristics for the single electron population case obtained in the present experiments using the axial LP placed on axis at $z = -4$ cm (inside the PSS), for hydrogen gas pressure ≈ 3 mTorr and microwave power ≈ 650 W. The red dashed line in the figure is *the linear fit to the deep, negatively biased portion* of the $I$ - $V$ plot. It is seen that the *$I - V$ plot lifts off from the linear fit* at the point ($I_s$, $V_s$), *indicating increasing contribution from a second species for $V > V_s$*. Here $I_s$ is the ion saturation current and $V_s$, the corresponding voltage. Since electrons are the only other species in the present case, one can see that with *reducing negative bias* there would be growing electron contribution to the probe current. Thus, ions and electrons both contribute for $V > V_s$, which marks onset of the *transition regime*.

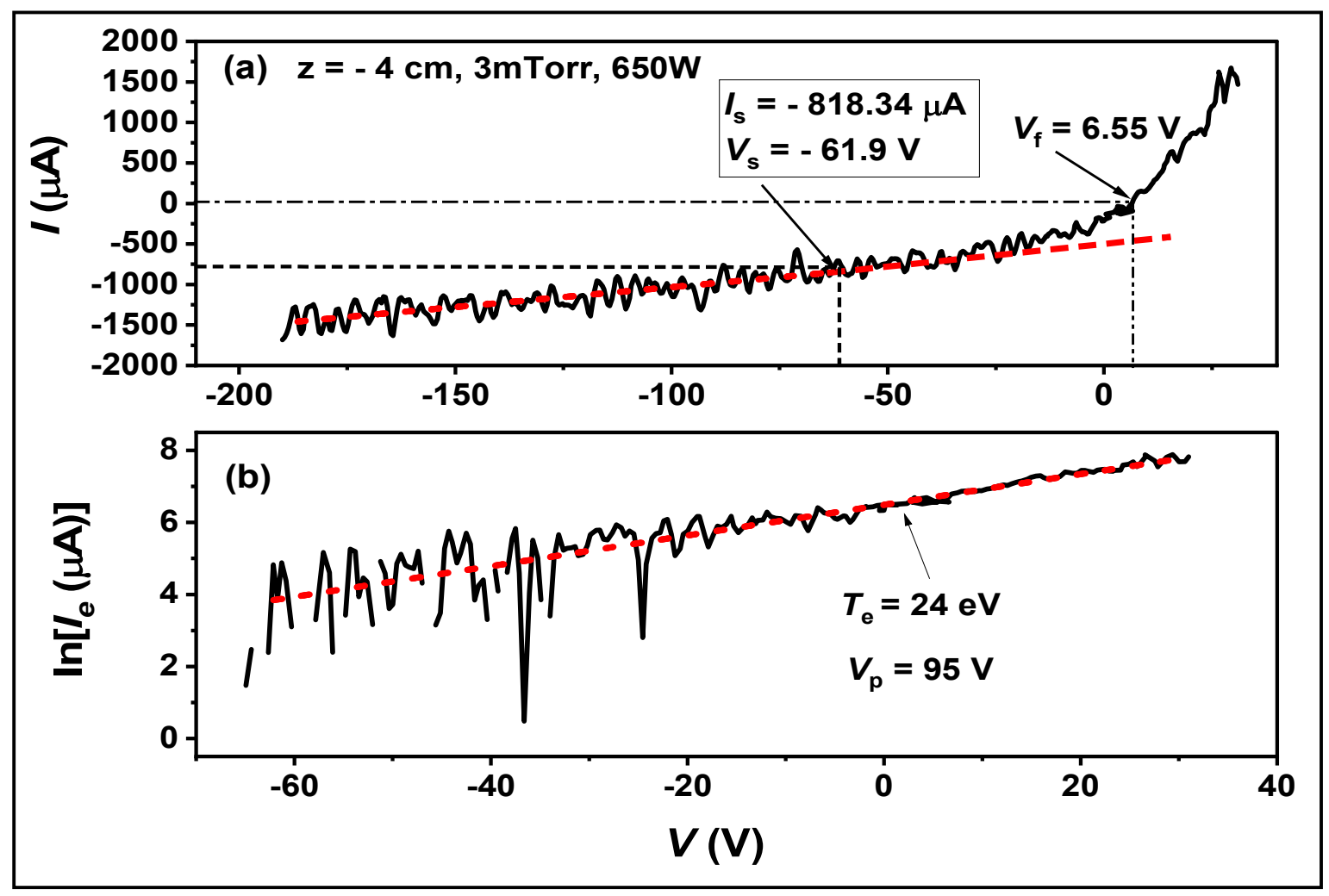

**FIG. 5. (a) Typical *I-V* characteristic of axial LP obtained at $z$ = - 4 cm (inside PSS) for hydrogen plasma at ~3 mTorr, 650 W. The red dashed line is the linear fit to the *I–V* plot at deep negative biases. The lift-off of the *I–V* plot from the linear fit gives the ion saturation current, $I_s$ = - 818.3 A at a bias voltage, $V_s$ = - 61.9 V. (b) Plot of ln ($I_e$) versus probe voltage, *V*. The red dashed line gives the linear fit. Inverse slope of the fit gives the electron temperature, $T_e$ = 24 eV.**

To determine the electron contribution $I_e$, one subtracts $I_s$ from the total probe current $I$ for $V > V_s$. It is generally assumed that electrons are in thermal equilibrium, and that their density obeys the Boltzmann relation, ~ exp $\{e (V - V_p)/T_e\}$; here, $V_p$ is the space or plasma potential. Thus, a plot of ln ($I_e$) versus $V$, as shown in Fig. 5(b), is expected to yield a straight line. The red dashed line in Fig. 5(b) gives the linear fit to the ln ($I_e$) versus $V$ plot, whose inverse slope gives $T_e$. In the present experiments, it was not possible to enter the electron saturation regime as the currents became too large, beyond limits of the LP power supply.

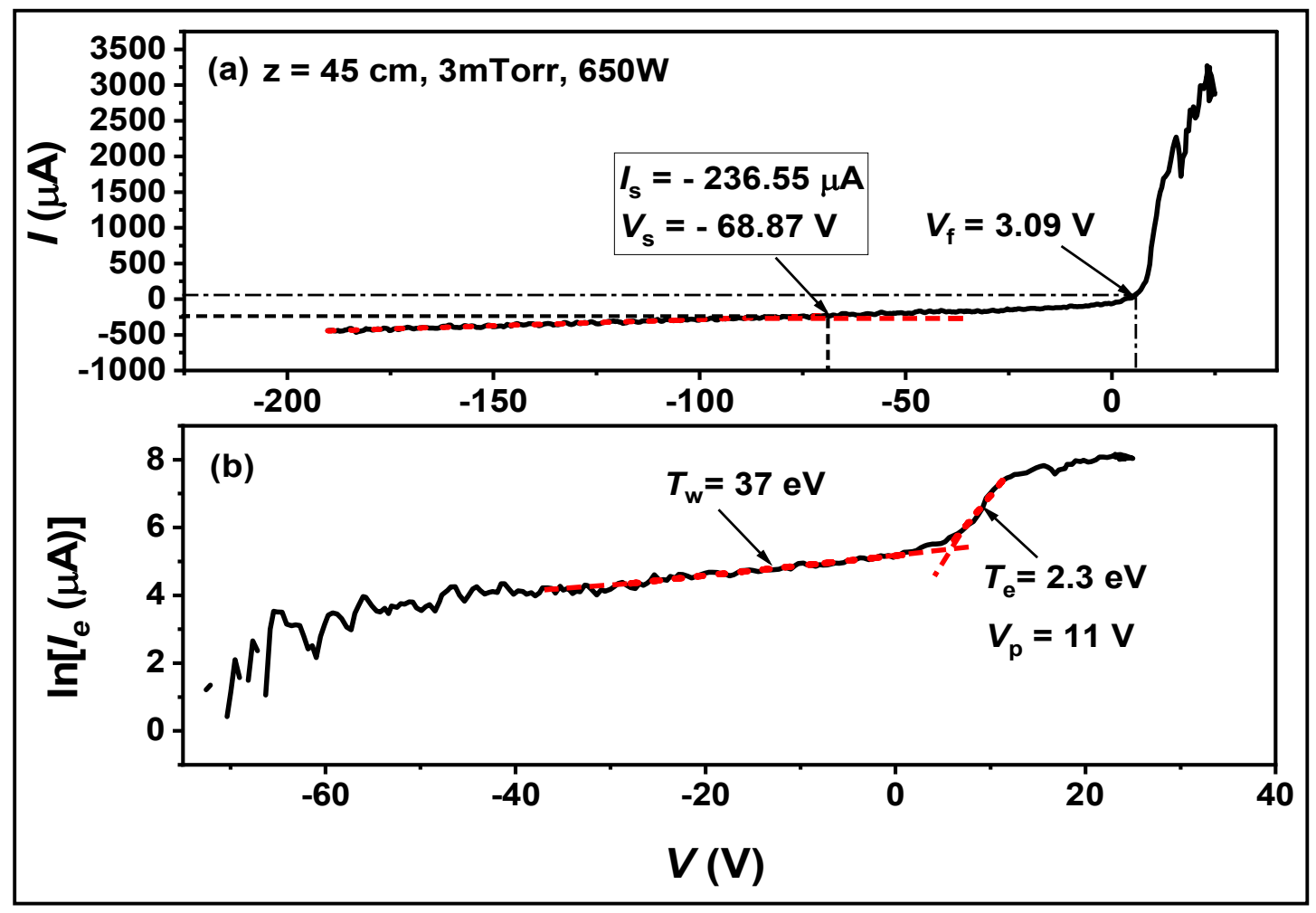

**FIG. 6. (a) Typical I-V characteristic of axial LP obtained at z = 45 cm (inside expansion chamber) for hydrogen plasma at ~3 mTorr, 650 W with similar analysis as in Fig. 5. $I_s$ = - 236.6 A and $V_s$ = - 68.9 V (b) ln ($I_e$) versus probe voltage, V, showing presence of two electron populations: $T_w$ = 37 eV (warm population) and $T_e$ = 2.3 eV (bulk population).**

***Two Electron Temperature Populations:*** A typical case of LP data with two electron populations is shown in Fig. 6. Fig. 6(a) gives the $I - V$ characteristics from which ($I_s$, $V_s$) are determined as for the single temperature case. Unlike Fig. 5(a), however, the rise in the probe current, $I$ beyond $V_s$ is *almost flat and very gradual*, which is indicative of a high temperature, low density electron population. Subtracting $I_s$ from $I$ for $V > V_s$ yields $I_e$. The plot of ln ($I_e$) versus $V$ is shown in Fig. 6(b). The plot is seen to be linear, indicating that the electron density obeys the Boltzmann relation. The red dashed curve is the linear fit to the ln ($I_e$) versus $V$ plot, and its inverse slope gives the temperature of the electrons, designated as *warm electrons*. One notes that the ln ($I_e$) plot *lifts off* sharply from the linear fit at low values of $V$ (≈ 0), which is indicative of the onset of a second, low temperature, high density (bulk) electron population. Following an identical procedure as above, it is possible to isolate the bulk electrons in the next step. The LP data were analysed on an interactive software (MATAB GUI) developed in-house at the Plasma Lab, IIT Delhi.[28, 29]

### 2.3 Retarding Field Energy Analyzer (RFEA)

The plasma from the PSS region expanding into the EC was monitored for high energy ion beams using a Retarding Field Energy Analyser (RFEA). A schematic of the RFEA for measuring ion energies on the axis of the EC and the PSS is shown in Fig. 7(a). It consists of a hollow, stainless steel (ss) cylindrical head of length ≈ 28 mm and dia ≈ 25 mm. Its plasma facing surface (normal to the chamber axis) has a ≈ 6 mm aperture at its centre that only permits ions close to the axis to enter the analyzer. It was found that the insulating material (teflon) inside the RFEA was severely damaged because of extreme heating caused by the bombarding ions. To avoid such damage, the ss body of the RFEA was encased in a graphite cup, which in turn was protected by a ceramic shield. The graphite prevents local heating of the analyzer body by absorption of any residual microwaves around the RFEA in addition to distributing the heat around the RFEA surface. The ceramic cup protects the graphite from direct bombardment by the ions, preventing sputtering and pitting of its surface. The RFEA uses only three ss grids: an electron repeller (ER), ion discriminator (ID) and an electron suppressor (ES) with a stainless-steel collector (C) plate behind the ES, for collecting all ions reaching it. Fig. 7(b) shows the voltages applied on each grid.

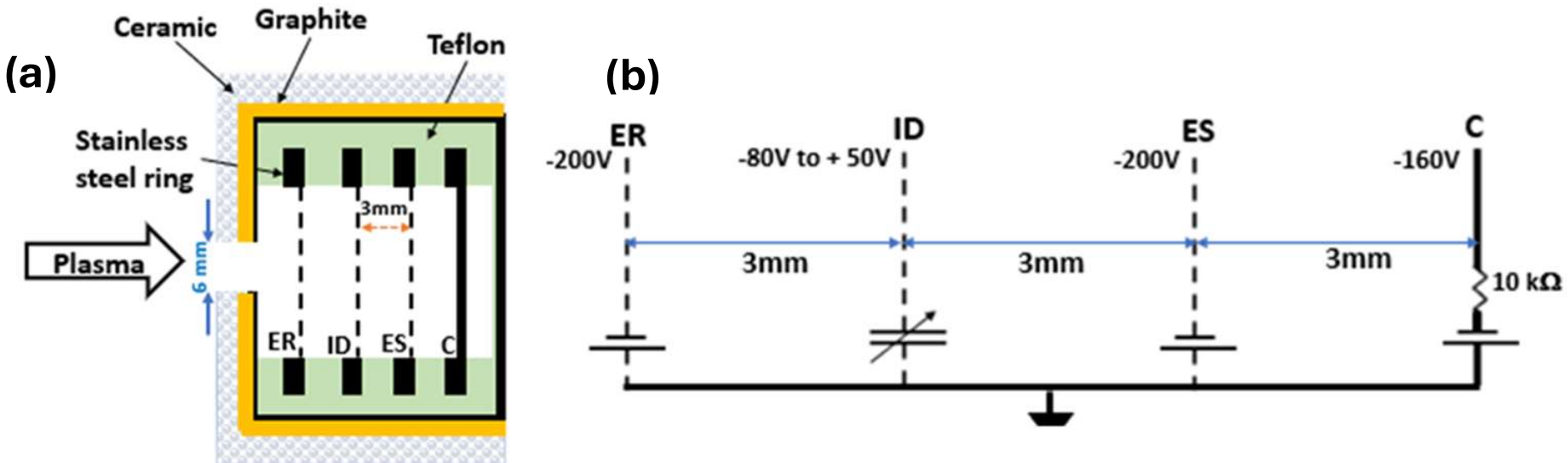

**FIG. 7. The Retarding Field Analyzer (RFEA): (a) schematic of the RFEA and (b) The grid biasing layout used in the RFEA where the acronyms signify ER: Electron Repeller, ID: Ion Discriminator, ES: Secondary Electron Suppressor, C: Collector**

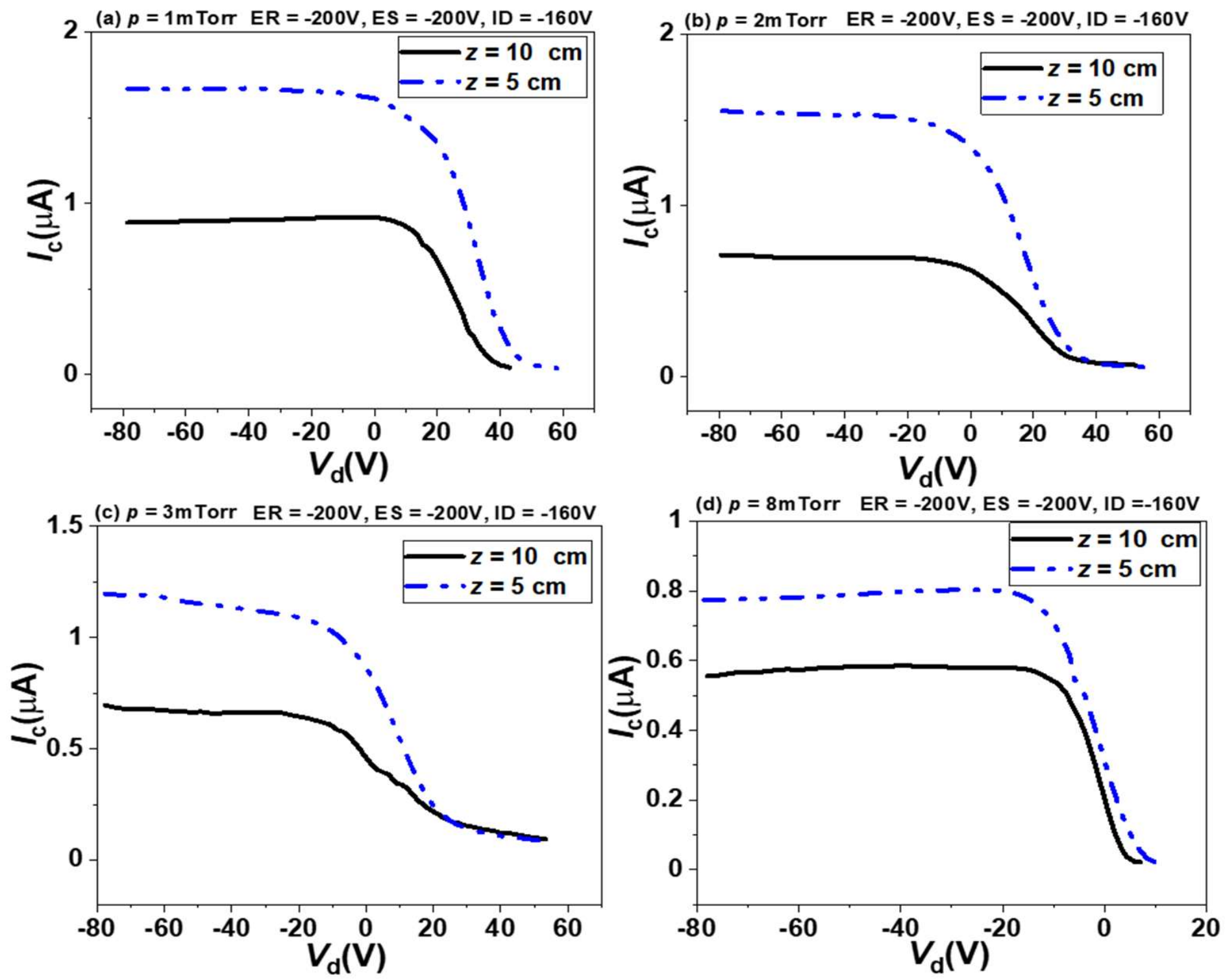

**FIG. 8.** $I_c$ **(**$V_d$**) characteristics measured using the RFEA positioned at** $z \approx 5$ **and ≈ 10 cm in front of the GEPS for (a) ≈ 1 mTorr; (b) ≈ 2 mTorr; (c) ≈ 3 mTorr and (d) ≈ 8 mTorr pressure and microwave power of 650 W.**

Due to the presence of highly energetic electrons in the plasma, the bias on ER had to be decreased to deep negative values ≈ -200 V before the measured ion current at ID could attain saturation. Furthermore, because a high energy (≥ 60 eV) ion beam was anticipated in the present case, the bias on ID was not varied from ≈ - 200 V (as would have been expected) but from ≈ - 80 V (retarding with respect to the ER). It also turned out that at a bias of about + 50 V the ion current was reduced to zero, even for a position ≈ 5 cm in front of PSS where the current is expected to be high.

It can therefore be seen that the voltage difference between the ER and ID reaches a maximum value ≈ 250 V. The voltage on the collector was ≈ -160 V. As seen from Fig. 7(b), a large voltage difference can also occur between ID and ES, as well ≈ 250 V. Under these conditions, to avoid arcing between adjacent grids, the inter-grid spacing was kept ≈ 3 mm, uniformly for all grids.

The ss body of the analyser was grounded via the wall of EC. Although the RFEA can disturb the plasma, the measurements were repeated with similar results, and no noticeable perturbations of the plasma were observed. Ion energy distribution functions were generated by differentiating the $I_c - V_d$ plots with respect to $V_d$ at the different axial locations.

The variations of the collector current $I_c$ with respect to the discriminator voltage $V_d$ were taken at $z \approx 5$ cm and 10 cm for pressure ranges from $p \approx 1$ - 8 mTorr at 650 W microwave power. Figs. 8 (a – d) represent typical profiles of the collector current $I_c$ versus the discriminator voltage $V_d$ at ≈ 1 mTorr, ≈ 2 mTorr, ≈ 3 mTorr and ≈ 8 mTorr pressures respectively, for an input microwave power 650 W, CW. One notes that the $I_c$ profiles are those of current beams. It is seen that on account of the deep negative voltages applied on the different grids, most of the change in $I_c$ occurs for $V_d$ in the range, ≈ –30 to +35 V. $V_d$ is measured with respect to ground, whereas one needs an estimate of the kinetic energy of the ions with respect to the plasma, where the reference potential is the plasma potential, $V_p$. Hence one has to change the reference from ground to $V_p$ which requires defining the variable, $E = V_d + V_p$, so that positive values of $E$ (i.e., $E > 0$) represent directly the kinetic energy of the ions in the plasma, with $E = 0$ (i.e., $V_d = -V_p$) corresponding to the zero of the kinetic energy. Plots of $f(E)$ versus $E$ have been presented in Figs. (12 – 13) using the appropriate values of $V_p$.[31, 32]

## 3 Plasma Characterisation Results and Analysis

Experiments were performed using hydrogen gas under different operating conditions at neutral gas pressure ranging from ≈ 1 to ≈ 8 mTorr and fixed microwave power ≈ 650 W.

### *3.1 Axial and Radial LP Profiles and RFEA Results*

Figs. 9 and 10 give axial profiles of the bulk plasma density $n$, bulk electron temperature $T_e$, the warm electron density and temperature, $n_w$, $T_w$ and the plasma potential $V_p$ for pressures, ≈ 1, ≈ 2, ≈ 3, ≈ 6 and ≈ 8 mTorr. Fig. 11 gives the radial LP profiles at the lowest (≈ 1 mTorr) and highest pressures (≈ 8 mTorr) at two axial locations. Figs. 12 and 13 give plots of the ion distribution function, $f(E)$.

***Low Pressure Axial Profiles – Density Jump and Double Layer*:** The resonantly heated electrons in the ECR zone escape rapidly leaving behind a plasma that is slightly positively charged. In steady state, the residual plasma acquires a high positive potential that gives rise to an ambipolar electric field, which not only arrests the rapid loss of electrons but *also ensures that both electrons and ions leave together with a common velocity towards the PSS exit* (at $z = 0$). It turns out, that *the hot electrons can initiate a second ionization zone*, that results in a *density jump* and formation of a *double layer* (DL), which play an important role in the ensuing plasma dynamics.

The *dashed red line* at $z = z_2$ in Figs. 9 (a – c) specifies the location of the *sharp density jump* (DJ), from $z = z_2^-$ to $z = z_2^+$. As noted above, such density jumps arise because of the *second ionization zone* (SIZ) that is initiated just before the DJ. The approximate start location of the SIZ is given by the *dashed blue line* at $z = z_1$ in Figs. 9. The gap, $\Delta z = z_2 - z_1$ also coincides with a large fall in $V_p$ (see figure), signalling the presence of a *double layer* (DL) there, so that the DL and the SIZ overlap. For example, at ≈ 1 mTorr pressure, $z_1 \approx -3$ cm, $z_2 \approx +2$ cm so that the width of the DL, $w_{DL} = z_2 - z_1 \approx 2 - (-3) = 5$ cm with the potential drop, $\Delta V_{DL} = V_p(z_1) - V_p(z_2) \approx 106 - 34 \approx 72$ V. In general, at low pressures, the potential drops across the DLs lie within an approximate range, $\Delta V_{DL} \approx 45 - 75$ V [varying from ≈ (90 – 106) V to about ≈ 30 – 49 V] with a width in the range, $w_{DL} \approx 1 - 5$ cm. In some cases (≈ 1 and 3 mTorr, most clearly) one also notices the onset of gradual falls in $T_e$ between $z_1$ and $z_2$, before $T_e$ experiences a sharp, step-like fall at $z_2$ from ≈ 25 – 30 eV to ≈ 10 – 15 eV. The falls in $T_e$ and $V_p$ may be attributed to the onset of the SIZ at $z_1$ since the latter implies a conversion of the hot electrons (that take part in ionization) into low energy electrons along with addition of low energy secondary electrons born from ionisation.

***Post DJ Low Pressure Axial LP Profiles in the Expansion Chamber*:** It is observed that for ≈ 1 – 3 mTorr, the bulk density *just after* the DJ is typically $n \approx (1 - 2) \times 10^{11}$ cm$^{-3}$, following which the density falls steeply in the first ≈ 7 – 10 cm or so in the expansion chamber (EC), before flattening out at ≈ $5 \times 10^{10}$ cm$^{-3}$.

The plasma formed at and after the DJ also shows several other features. Following its sharp fall at the DJ, $T_e$ falls relatively slowly to stabilise at ≈ 5 eV in the rest of the EC. Likewise, following the DL, $V_p$ decreases smoothly to ≈ 10 – 12 V in the first ≈ 7 – 10 cm or so, flattening out thereafter. Just after the DJ, one also observes for all pressures (≈ 1 – 3 mTorr), the sudden birth of a warm electron population with density $n_w \approx (4 - 9) \times 10^9$ cm$^{-3}$, that settles down quickly to ≈ $1 \times 10^9$ cm$^{-3}$ along the EC. Typically, $\frac{n}{n_w} \approx 10 - 25$ within the EC. The warm electron temperature, $T_w \approx 40 - 60$ eV is nearly uniform over the entire chamber. It is interesting to note the *complete absence of such a population prior to the DJ* (see Section 3.3).

***High Pressure Axial LP Profiles in PSS and EC*:** The high-pressure (≈ 6, 8 mTorr) LP data are displayed in Figs. 10. With increasing pressure both, $z_1$ and $z_2$, and hence, the SIZs, DLs and DJs all move *inwards*, from outside the PSS to inside it. In fact, at pressures ≈ 6, ≈ 8 mTorr, the SIZ / DJ move so deep into the PSS (see Figs. 10) that these are no longer accessible to the LP. This is a consequence of *shrinking ionization mean free paths with increasing pressure* that pull the SIZs inwards, closer to the ECR zone. The warm electrons are also present at these pressures with both, $n$ and $n_w$ being a little higher. For both pressures $n$ falls sharply within the first ≈ 5 cm or so, becoming fairly flat and settling at ≈ $10^{11}$ cm$^{-3}$ thereafter. $n_w$ however, displays a profile like that of $n$ at the lower pressure settling quickly at a low value (≈ $10^8$ cm$^{-3}$); its drop at the higher pressure is gentler falling to ≈ $5 \times 10^8$ cm$^{-3}$ near the chamber end.

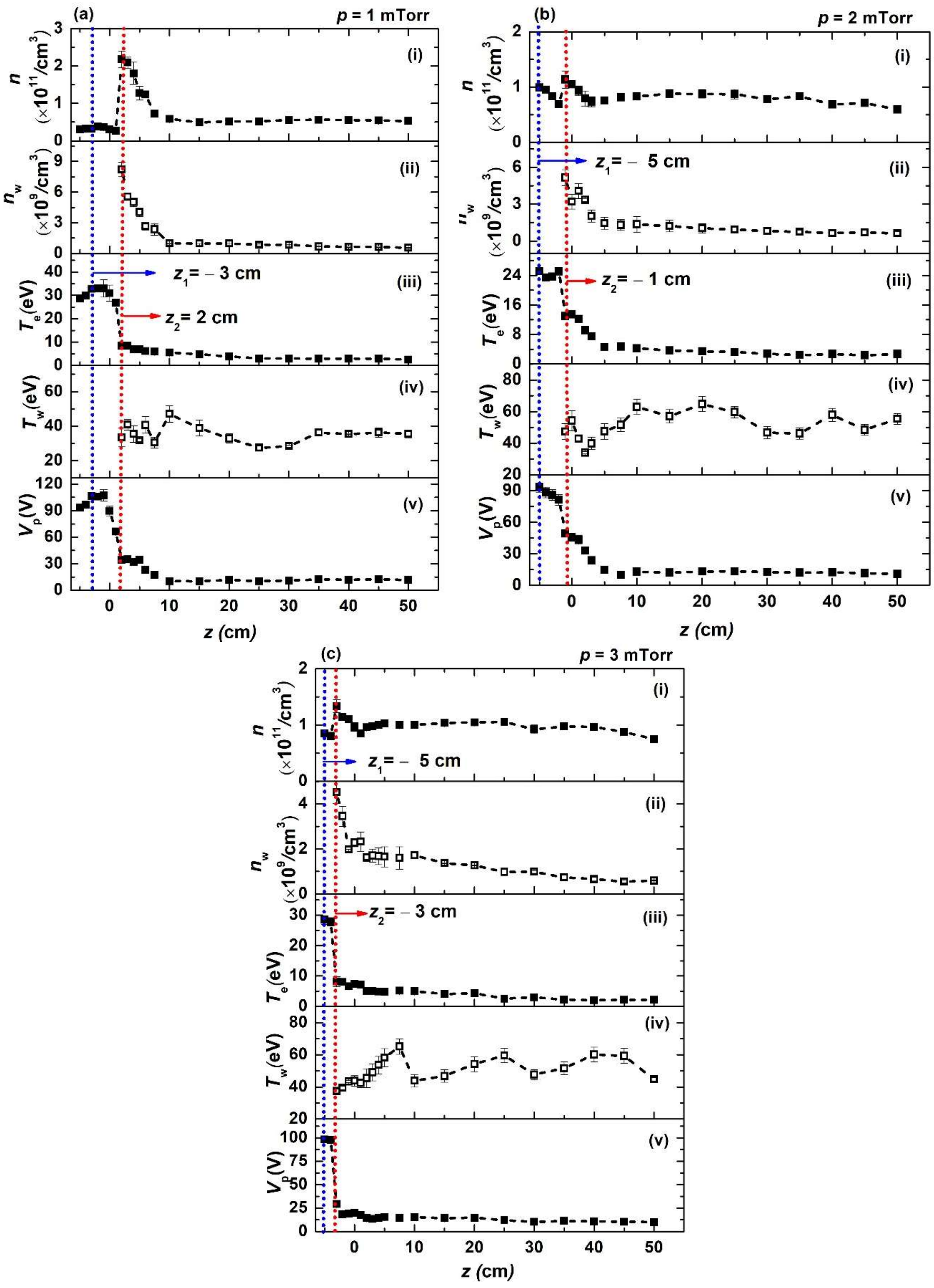

**FIG. 9.** Axial profile of hydrogen plasma parameters obtained using ≈ 650 W microwave input power at pressures: (a) ≈ 1 mTorr; (b) ≈ 2 mTorr; (c) ≈ 3 mTorr. The plasma parameters in each subplot are (i) bulk electron density ($n$); (ii) warm electron density ($n_w$); (iii) bulk electron temperature ($T_e$); (iv) warm electron temperature ($T_w$) and (v) plasma potential ($V_p$). The blue and red dashed lines at $z = z_1$ and $z = z_2$ respectively, give approximate locations of onset and termination of the SIZ and DL.

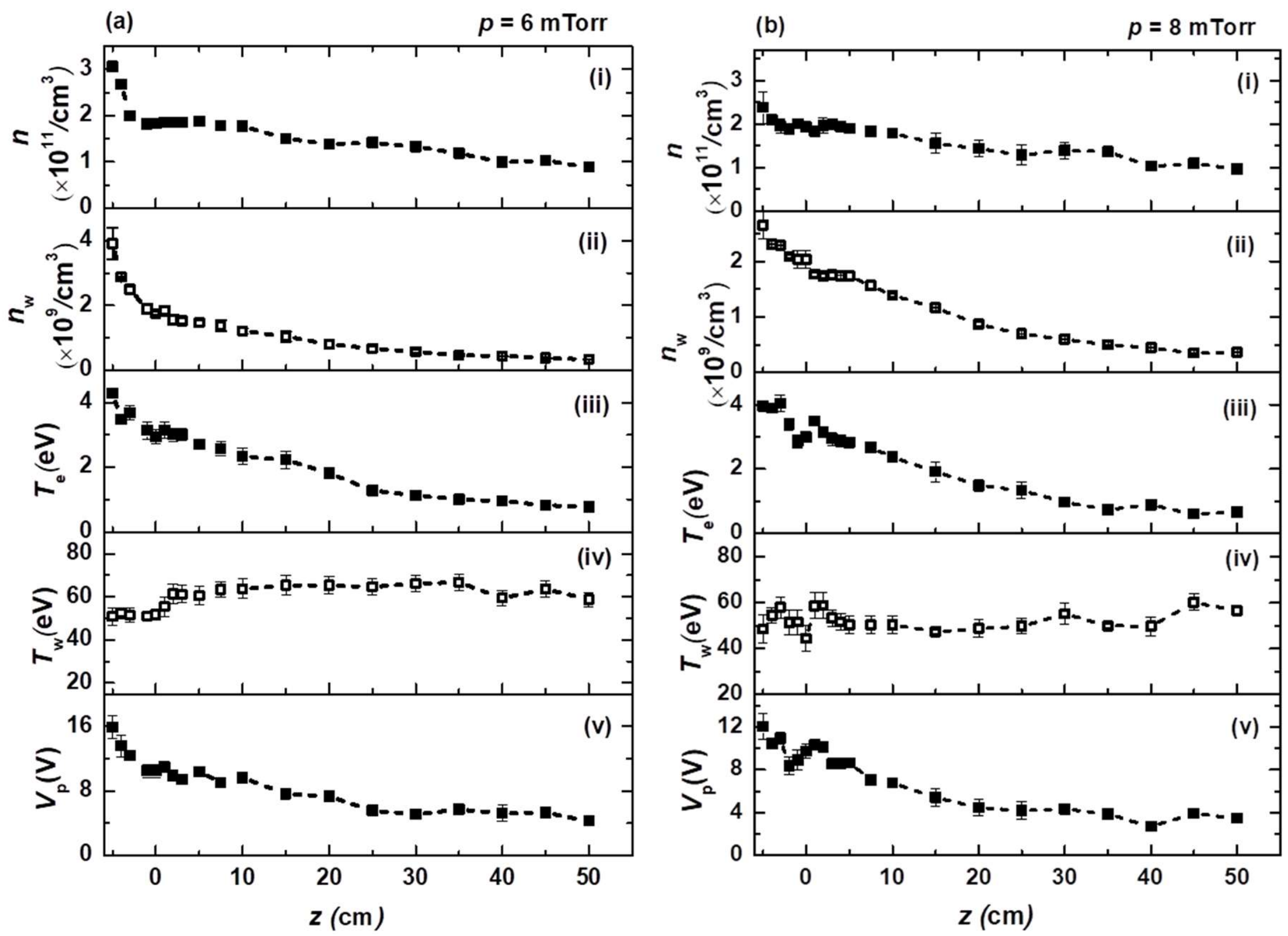

**FIG. 10. Similar figure as in Fig. 9 for pressures: (a) ≈ 6 mTorr; (b) ≈ 8 mTorr.**

Although $T_e$ is ≈ 4 eV after the SIZ, it falls to fairly low values ≈ 0.5 – 1 eV towards the chamber end, signalling cool electrons that can participate in negative hydrogen ion production. $T_w$ ≈ 40 eV throughout the EC.

***Radial LP Profiles***: Radial LP profiles are presented for ≈ 1mTorr and ≈ 8 mTorr in Figs. 11 (a) and (b) respectively, at axial locations $R_1$ ($z$ ≈ 13.8 cm) and $R_2$ ($z$ ≈ 38.8 cm). For $p$ ≈ 1mTorr, at location $R_1$ ($z$ ≈ 13.8 cm) the on-axis ($r$ ≈ 0) parameters are $n \approx 0.47 \times 10^{11}$ cm$^{-3}$, $T_e$ ≈ 4.23 eV, $n_w \approx 0.66 \times 10^9$ cm$^{-3}$, $T_w$ ≈ 47 eV and $V_p$ ≈ 14 V. At this location, one sees a slight decrease in $n$ and $V_p$ with the radius, although $T_e$ and $n_w$ are nearly uniform. The decrease in $T_w$ however is seen to be quite pronounced. It can be seen that because $T_w$ falls from ≈ 50 eV to ≈ 20 eV ionization by the warm electrons would be more sensitive to $T_w$ than $n_w$. On the other hand, the $n$ and $n_w$ profiles in Fig. 11 (b) are seen to be practically identical at $R_1$, indicating that plasma production at ≈ 8 mTorr is controlled by the warm electron density rather than $T_w$ (≈ 47 eV), since $n_w$ and not $T_w$ would be the controlling factor for ionization.

The $T_e$ profile resembles that of $n$, dropping from ≈ 1.5 eV near the axis to about ≈ 0.6 eV at $r$ ≈ 6.5 cm. These results indicate that plasma reaches the off-axis regions (≤ 6 – 10 cm) mainly by ionization by the warm electrons and not due to plasma arriving from the source mouth.

At $R_2$ all plasma parameters (except $T_e$) are seen to be fairly uniform with the radius at both pressures. While $T_e$ is reasonably uniform at the lower pressure in Fig. 11(a), at the higher pressure (≈ 8mTorr) it decreases from ≈ 1.5 eV near the axis to about ≈ 0.6 eV at at $r$ ≈ 6.5 cm.

This is quite attractive from the viewpoint of negative hydrogen ion production, which requires very low temperature bulk electrons to prevent dissociation of the $H^-$ ions.

***RFEA Data***: As already discussed in Section 2.2, the $I_c – V_d$ data in Figs. 8 indicate the presence of ion beams. The ion energy distribution functions displayed in Figs. 12 and 13 were obtained by differentiating $I_c$ with respect to $V_d$ and changing variables from the ion discriminator voltage $V_d$ to the kinetic energy $E$ of the ions in the plasma. These indicate high ion energies, the energy decreasing monotonically with the pressure. Typically, at ≈ 1 mTorr (Fig. 12 a) one finds ions with *peak* energy, $E_b$ ≈ 63 eV with spread $\Delta E_b$ ≈ 22 eV, at $z$ ≈ 5 cm and $E_b$ ≈ 37 eV, $\Delta E_b$ ≈17 eV at $z$ ≈ 10 cm. At $p$ ≈ 2 mTorr, $E_b$ decreases to ≈ 30 eV ($\Delta E_b$ ≈ 23 eV) at $z$ ≈ 5 cm with $E_b$ ≈ 29 eV ($\Delta E_b$ ≈ 24 eV) at $z$ ≈ 10 cm (Fig. 12 b). The corresponding values of $V_p$ are also indicated for each case. Similarly, one may read off the values for $p$ ≈ 3 and 8 mTorr from Fig. 13.

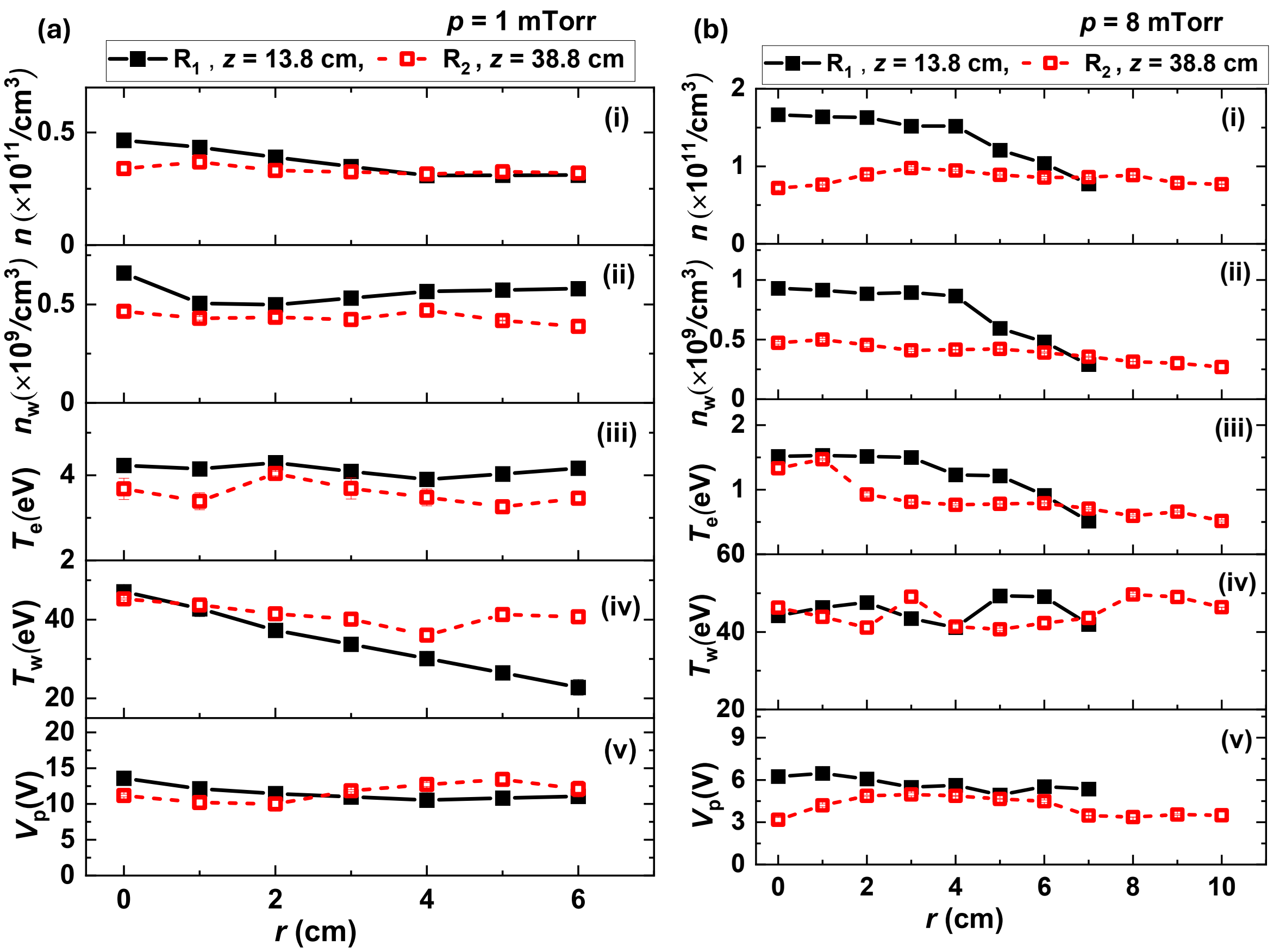

**FIG. 11. Radial profiles of hydrogen plasma parameters at different pressures using 650 W microwave power at axial locations $R_1$ and $R_2$, corresponding to $z \approx 13.8$ cm and $\approx 38.8$ cm respectively. (a) ≈ 1 mTorr (b) ≈ 8 mTorr; (i) bulk electron density ($n$); (ii) warm electron density ($n_w$); (iii) bulk electron temperature ($T_e$); (iv) warm electron temperature ($T_w$) and (v) plasma potential ($V_p$).**

## *3.2 Analysis of Axial Plasma Profiles*

***Formation of* SIZ *and* DJ**: The presence of the SIZ / DJ discussed above implies that the distance of the ECR zone (at $z \approx -9.1$ cm) to the location of the SIZ must be of the order of the *ionization mean free path* (= $\lambda_{iz}$) at that pressure. It must be remembered however, that while the thermal speed of the hot electrons determines the ionization frequency $\nu_{iz}$, ionization mean free paths computed using thermal speeds would be much higher than those encountered in the present work (≈ 4 – 6 cm). On the other hand, *inside the PSS it is not the random thermal motion that is of relevance*, *but the flow velocity* with which the electrons proceed along the PSS. It was stated above that electrons and ions must have a common ambipolar flow velocity to meet the requirements of quasineutrality. The latter velocity is much smaller than the thermal speed of electrons. *This retardation of the electrons is achieved because of slowing down by the ambipolar electric field and due to friction from the frequent elastic collisions with neutrals*. In steady state, the electrons are slowed down to the flow velocity of the ions.

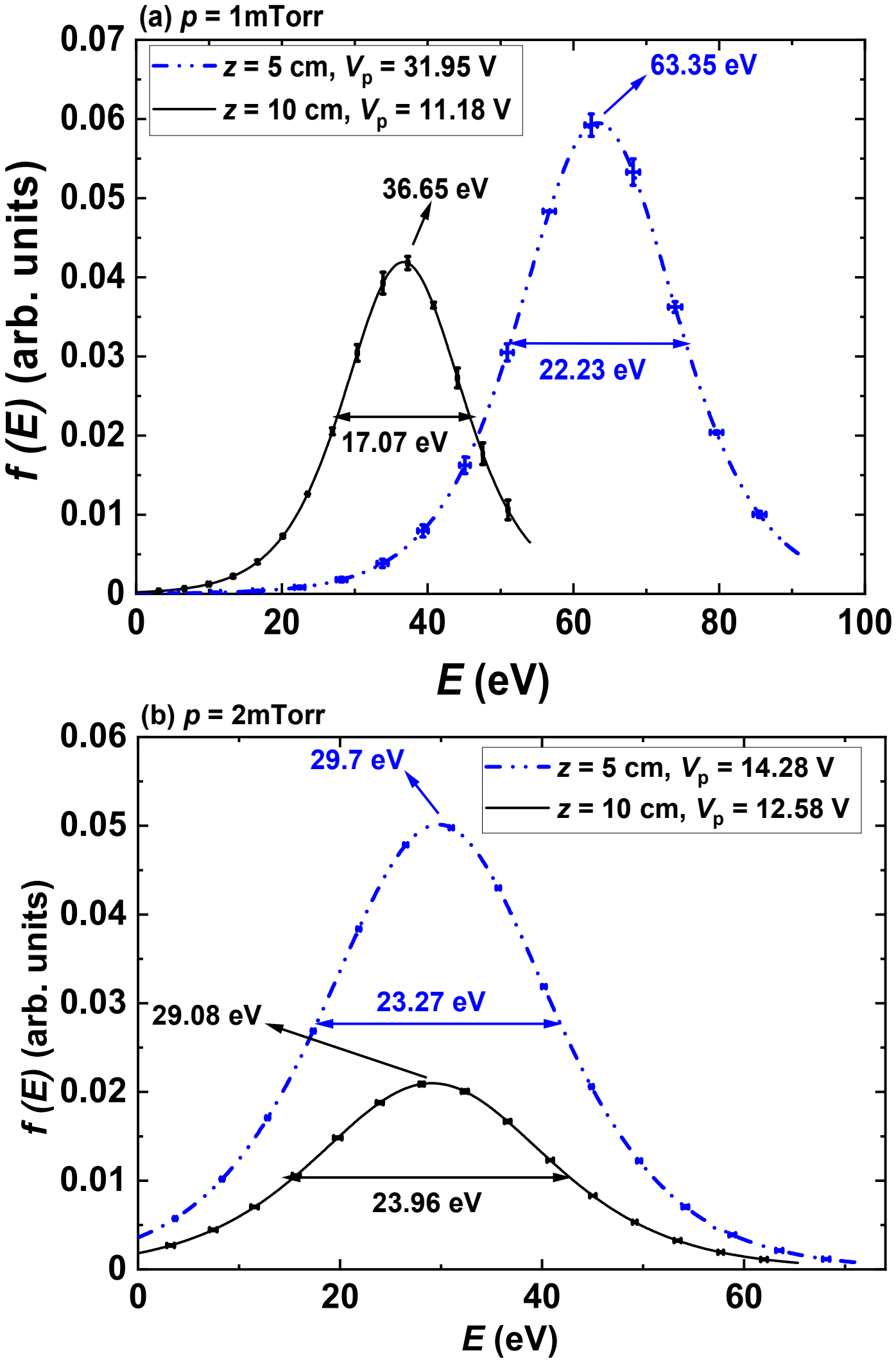

**FIG. 12. The ion energy distribution function *f* (*E*) with respect to *E* derived from the RFEA characteristics given in Fig. 8. (a) ≈ 1 mTorr; (b) ≈ 2 mTorr pressure and microwave power of ≈ 650 W**

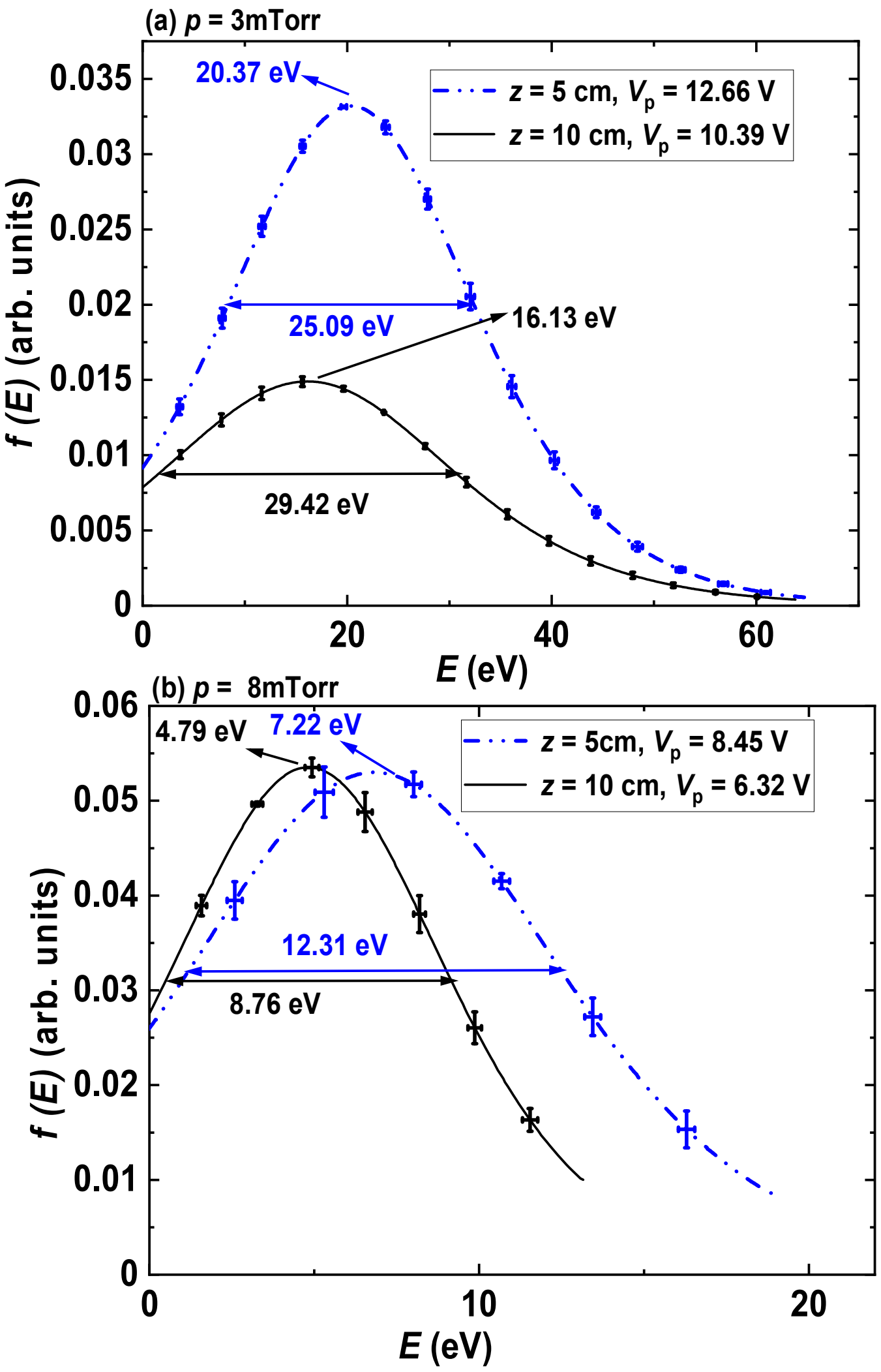

**FIG. 13. The ion energy distribution function f (*E*) with respect to *E* derived from the RFEA characteristics given in Fig. 8. (a) ≈ 3 mTorr; (b) ≈ 8 mTorr pressure and microwave power of ≈ 650 W.**

In light of the above therefore, the ionization mean free path, $\lambda_{iz}$ may be expressed as $\lambda_{iz} = \frac{\mathrm{v}}{\nu_{iz}}$, where v is the *flow* velocity of the *electrons and ions within the PSS*. The velocity v can be determined by solving the *ion flow equation* in the paraxial region between the ECR zone and the SIZ, in the presence of the steady state ambipolar electric field E and friction due to elastic and charge exchange collisions with the neutrals. Also, since the GEPS magnetic field is nearly parallel to the PSS axis (in its paraxial region), it will have little effect on the axial ion motion and may be ignored. In steady state, the velocity v obeys $\frac{\partial \mathrm{v}}{\partial t} \approx 0$, so that the momentum equation for a cold (≈ zero pressure) ion fluid reads,

$$\mathrm{v}\frac{\mathrm{dv}}{\mathrm{d}s} = \frac{e\mathrm{E}}{M} - \nu_{\mathrm{mi}}\mathrm{v} \qquad (1)$$

In (1), the density *n* has cancelled out and *e*, *M* are the ion charge and mass. *s* is the local *axial* coordinate *along the axis* with its origin ($s = 0$) at the ECR zone at $z \approx -9.1$ cm, giving the relation, $s(z) = z + 9.1$ cm, in general. Eq. (1) was solved up to $s = s_1$, *the onset of the SIZ* or $z = z_1$ in Fig. 9, giving, $s_1 = s(z_1) = z_1 + 9.1$ cm. The momentum transfer collision frequency, $\nu_{\mathrm{mi}} = n_{\mathrm{g}}\sigma_{\mathrm{tot}}\mathrm{v}$, where the total cross section $\sigma_{\mathrm{tot}} = \sigma_{\mathrm{el}} + \sigma_{\mathrm{ce}}$. $\sigma_{\mathrm{el}}$, the *elastic* scattering cross section is given by the Langevin formula as,[34] $\sigma_{\mathrm{el}} = \frac{C}{\mathrm{v}}$, where the constant $C = 2.7 \times 10^{-1}\ \frac{\mathrm{m}^3}{\mathrm{s}}$ for hydrogen and the charge exchange cross section, $\sigma_{\mathrm{ce}} \approx 3.2 \times 10^{-1}\ \mathrm{m}^2$ is approximately a constant in the energy range of concern here; $n_{\mathrm{g}}$ is the gas density. Lack of sufficient data in the pre-SIZ region ($0 < s < s_1$) in Fig. 9, does not permit straightforward evaluation of the ambipolar electric field E, *inside the PSS*. However, it was noted earlier that the *potential drop* from the the ECR zone to the SIZ (i.e., across the PSS) and the ensuing ambipolar electric field have to be large enough to trap an *average* electron. Since the average electron has energy ~ $T_{\mathrm{e}}$, one may set the potential drop from the ECR zone to the SIZ ≡ $\Delta V_{\mathrm{PSS}} \sim T_{\mathrm{e}}$, so that, $\mathrm{E} \approx \frac{\Delta V_{\mathrm{PSS}}}{s_1} \approx \frac{T_e}{s_1}$. Here, $T_{\mathrm{e}}$ is the average temperature of the electrons in the PSS (assumed constant).

To solve (1), one sets the initial velocity $\mathrm{v} = 0$ at $s = 0$, the ECR location on the axis. It turns out that it is more convenient to solve for $\epsilon$, the kinetic energy of an ion with flow velocity v, i.e., use $\epsilon = \frac{\frac{1}{2}M\mathrm{v}^2}{e}$ instead of v. One has therefore,

$$\frac{d\epsilon}{ds} = \mathrm{E} - n_g\left\{\left[\frac{2M}{e}\right]^{0.5} C\sqrt{\epsilon} + 2\sigma_{\mathrm{ce}}\epsilon\right\} \qquad (2)$$

Eq. (2) was solved from $s = 0$ to $s = s_1$ using the initial condition $\epsilon = 0$ at $s = 0$. The electric field E was calculated for each pressure as per the ansatz discussed above. The distance $s_1$ from the ECR zone to the SIZ is found from Figs. 9 for each pressure. Eq. (2) is nonlinear and was solved numerically for the ion energy $\epsilon(s)$ inside the PSS. As discussed in detail earlier, to preserve quasineutrality the electrons in the PSS *must also have the velocity* $\mathrm{v}(s)$ and so arrive at the SIZ with velocity $\mathrm{v}(s_1)$. Since the electron velocity *varies* along the PSS, so does $\lambda_{\mathrm{iz}}$ $[= \frac{\mathrm{v}}{\nu_{\mathrm{iz}}}]$. On the other hand to determine $\lambda_{\mathrm{iz}}$ uniquely (for a given pressure) and *to refer all the calculated* $\lambda_{\mathrm{iz}}$ *to a common starting point* ($s = 0$), *one needs to ensure that all the electrons start from* $s = 0$ *with identical velocity, which in the present case may be taken to be the average velocity* $\mathrm{v}_{\mathrm{av}}$ *of the electrons / ions inside the PSS*. Thus, one may write $\lambda_{\mathrm{iz}} = \frac{\mathrm{v}_{\mathrm{av}}}{\nu_{\mathrm{iz}}}$ and determine $\mathrm{v}_{\mathrm{av}}$ using,

$$\mathrm{v}_{\mathrm{av}} = \left(\frac{1}{s_1}\right)\left[\frac{2e}{M}\right]^{0.5} \int_0^{s_1} \epsilon^{0.5}\, ds$$

Calculations were carried out for data presented in Figs. 9 for the three pressures, ≈ 1, ≈ 2 and ≈ 3 mTorr. The results are presented in Table 1. It lists $s_1$, the location of the SIZ, the temperature $T_e$ of the hot electrons, the ambipolar electric field E, the ion energy $\epsilon(s_1)$, the velocity $\mathrm{v}(s_1)$, the average velocity $\mathrm{v}_{\mathrm{av}}$, the ionization frequency $\nu_{\mathrm{iz}}$ and the mean free path, $\lambda_{\mathrm{iz}}$. $\nu_{\mathrm{iz}}$ was found using data given in,[34] and $\lambda_{\mathrm{iz}}$ was determined using $\mathrm{v}_{\mathrm{av}}$. One notes immediately that for all pressures, $\lambda_{\mathrm{iz}}$ compares favourably with the corresponding values of $s_1$. In fact, the $\lambda_{\mathrm{iz}}$ values are a little lower than the corresponding $s_1$ values.

**Table 1. Values of the ionization mean free paths ($\lambda_{iz}$) at ≈ 1, 2 and 3 mTorr for hot electrons, computed using the average velocity $v_{av}$ of the electrons in the PSS. The $\lambda_{iz}$ values may be compared with $s_1$ (= $z_1$ + 9.1) the approximate distance of the SIZ from the ECR zone ($s$ = 0 or $z$ – 9.1 cm), for each pressure. E is the electric field in the PSS.**

| Sr. No | $p$ (mTorr) | $s_1$ (cm) | $T_e$ (eV) | E = $T_e$ / $s_1$ (V/cm) | $\epsilon(s_1)$ (eV) | v($s_1$) (m/s) | $v_{av}$ (m/s) | $\nu_{iz}$ ($s^{-1}$) | $\lambda_{iz}$ (cm) = $v_{av}/\nu_{iz}$ |
|---|---|---|---|---|---|---|---|---|---|
| 1 | ≈ 1 | 6.1 ($z_1$ = -3) | 30 | 4.9 | 15.72 | $5.35\times10^4$ | $4.02\times10^4$ | $9.6\times10^5$ | 4.2 |
| 2 | ≈ 2 | 4.1 ($z_1$ = - 5) | 25 | 6.1 | 10.77 | $4.43\times10^4$ | $3.44\times10^4$ | $1.7\times10^6$ | 2.0 |
| 3 | ≈ 3 | 4.1 ($z_1$ = - 5) | 28 | 6.8 | 8.71 | $3.98\times10^4$ | $3.26\times10^4$ | $2.7\times10^6$ | 1.2 |

Despite limited knowledge of the PSS plasma in the GEPS, the above analysis clearly shows how an SIZ may form in such systems to give a boost to the bulk plasma density.

***Double Layer (DL) Formation***: Reviewing the above facts, one sees that in the neighbourhood of the SIZ there are actually two distinct plasmas with very different characteristics: A plasma P1 ($z \leq z_1$ or $s \leq s_1$) that flows in from the ECR zone and the plasma P2 ($z \geq z_2$) that is born due to ionization by the hot electrons of P1 and separated from the latter by the DL. In what follows, quantities referring to P1 will have the subscript 1, while those referring to P2, subscript 2. As seen above, P1 is characterized by moderately high densities ($n_1 \approx 1 - 5 \times 10^{10}$ cm$^{-3}$) hot bulk electrons ($T_{e1} \approx 30$ eV) and high plasma potentials with respect to the chamber walls ($V_{p1} \approx 90$ – 110 V). P2 on the other hand, has higher bulk plasma densities ($n_2 \approx 1 - 2 \times 10^{11}$ cm$^{-3}$), lower bulk electron temperatures ($T_{e2} \approx 10 - 15$ eV) and lower plasma potentials ($V_{p2} \approx 25 - 45$ V).

Based on the above, one may envision the following scenario taking place at the SIZ: The hot electrons of P1, after losing a considerable fraction of their energy in ionization, *mingle with the electrons born from the ionization to form the bulk electron population of* P2. As a result, P2 has an excess of negative charge (since the ionization process produces equal positive and negative charges). The extra charge and the lower $T_{e2}$ help lower $V_{p2}$ and keep it low. On the other hand, P1 on account of losing the electrons that take part in the ionization, is left with an excess of positive charge that helps increase $V_{p1}$. It is this constant supply of *opposite* charges to the two sides of the SIZ that helps generate and maintain a stable DL in steady state.

### *3.3 Generation of the Warm Electrons*

The mechanism for formation of the warm electrons is discussed in this section. It was originally developed by the authors for their work on hydrogen plasma generation using the CEPS,[31] and is presented here for completeness and illustration.

It is useful to go over the nomenclature introduced in an earlier section. The plasma flowing in from the ECR zone was termed P1, while that born due to ionization at the SIZ by the hot electrons of P1 and *separated from it by the DL* was labeled P2. Quantities referring to P1 were to have the subscript 1, while those referring to P2, subscript 2. As seen from the LP data, P1 is characterized by moderately high densities, hot bulk electrons ($T_{e1} \approx 30$ eV) and high plasma potentials ($V_{p1} \approx 90 – 110$ V) while P2 has higher bulk plasma densities than P1, lower bulk electron temperatures ($T_{e2} \approx 10 – 15$ eV) and lower plasma potentials ($V_{p2} \approx 25 –$ 45 V). As pointed out earlier, a distinct and separate warm electron population $P_{we}$ is also formed just *after* the DJ. It has a low density ($= n_w$) with $\frac{n_2}{n_w} \approx 10 - 25$ and a distinct warm temperature, $T_w \approx 40 – 60$ eV.

Now, the high energy electrons of P1 can overcome the potential barrier posed by the DL at $z = z_1$ (Fig. 9) to crossover to the other side of the DL at $z = z_2$ and one may use the Boltzmann relation to evaluate the fraction of electrons that cross over. However, this would not explain why the electrons that cross over do not merge with the bulk population of P2 (with temperature $T_{e2}$) or as to why they form a totally *separate* population with temperature $T_w$, *different* from the temperature $T_{e1}$ of the electrons in P1 from which they originated. It turns out actually, that in order to apply the Boltzmann relation to a population of particles in a certain region of a potential field, the population must be in thermal equilibrium in that region of the potential, meaning that its velocity / energy distribution function must not be *distorted* or *severely non-maxwellian* there. The latter statement is implicit in the derivation of the Boltzmann relation. In the present situation for instance, *bulk of the electrons of* $P_1$ *lose a* *considerable fraction of their energy in ionization at the DL*, which *heavily distorts their distribution function or "punches a hole in it" and takes it far away from thermal equilibrium.* Consequently, the Boltzmann relation can no longer be applied to the electrons of $P_1$ *at the DL*.

In the above scenario electrons of $P_1$ with *energy greater than* $\Delta V_{DL}$ *can nonetheless, cross over to the other side of the DL*. Let the electrons that can cross over, labeled $P_{DL}$ here, have density $n_{DL}$. In this context, it is important to note that *once the distribution function of the electrons in* $P_1$ *is distorted or fractured* (i.e., has become *non-maxwellian*), the electrons of $P_{DL}$ also separate out, *forming an independent group away from thermal equilibrium*. In *steady state*, however, these electrons of $P_{DL}$ will come to equilibrium (within themselves) forming a *new thermalized population,* $P_{we1}$ *with temperature* $T_{w1}$ that can exist on *both* sides of the DL. Thus, one may apply the Boltzmann relation to the population, $P_{we1}$ and estimate its density *after* the DL.

The *average* energy $< \epsilon_{\mathrm{PD}} >$ of an electron in the population $P_{DL}$ (with energy greater than $\Delta V_{DL}$) has to be determined from the electron population of $P_1$, *prior to its distribution function being disturbed* by the ionization process. It is given as, $< \epsilon_{\mathrm{PD}} > = \frac{E_{\mathrm{T}}}{F}$. Here, $F$ is the fraction of electrons with energy, greater than $\Delta V_{DL}$ and $E_T$, the *total, cumulative energy* of this fraction. Using the substitution, $\epsilon_{\mathrm{PD}} = \frac{m \mathrm{v}_{\mathrm{e}}^2}{2\mathrm{e}T_{\mathrm{e1}}}$ (here $v_e$ is the speed of the electrons and $m$ their mass) in the Maxwell-Boltzmann distribution and integrating over $\epsilon_{\mathrm{PD}}$ (instead of $v_e$) from $\frac{\Delta \mathrm{V_{DL}}}{\mathrm{T_{e1}}}$ to $\infty$, obtains,

$$F = \frac{2}{\sqrt{\pi}} \int_{\frac{\Delta V_{\mathrm{DL}}}{\mathrm{T_{e1}}}}^{\infty} \exp[-\epsilon_{\mathrm{PD}}] \sqrt{\epsilon_{\mathrm{PD}}}\, d\epsilon_{\mathrm{PD}}$$

$$E_{\mathrm{T}} = \left[\frac{2T_{\mathrm{e1}}}{\sqrt{\pi}}\right] \int_{\frac{\Delta V_{\mathrm{DL}}}{\mathrm{T_{e1}}}}^{\infty} \exp[-\epsilon_{\mathrm{PD}}]\, {\epsilon_{\mathrm{PD}}}^{3/2}\, d\epsilon_{\mathrm{PD}}$$

The above integrals can be expressed in terms of the *upper incomplete* Γ *function* (Abramowitz and Stegun, Chapter 6,[35]) defined as:

$$\Gamma(a,x) = \int_{x}^{\infty} \exp[-\epsilon]\,\epsilon^{a-1}\,d\epsilon$$

So that,

$$F = \left[\frac{2}{\sqrt{\pi}}\right]\Gamma\left(\frac{3}{2},\frac{\Delta V_{\rm DL}}{T_{\rm e1}}\right) \text{ and } E_{\rm T} = \left[\frac{2T_{\rm e1}}{\sqrt{\pi}}\right]\Gamma\left(\frac{5}{2},\frac{\Delta V_{\rm DL}}{T_{\rm e1}}\right)$$

With $F$ and $E_{\rm T}$ known, the electron density in $P_{\rm DL}$, $n_{\rm DL} \simeq F n_1$ and $<\epsilon_{\rm PD}> = \frac{E_{\rm T}}{F}$. In steady state, $P_{\rm DL}$ equilibrates to the warm electron population $P_{\rm we1}$ with temperature $T_{\rm w1}$ and density $n_{\rm w1}$. $T_{\rm w1}$ is obtained from $<\epsilon_{\rm PD}>$ by setting, (3/2) $T_{\rm w1} = <\epsilon_{\rm PD}>$, while $n_{\rm w1} \simeq n_{\rm DL}$. Since $P_{\rm we1}$ is in thermal equilibrium, one may apply the Boltzmann relation to determine $n_{\rm w2}$ the warm electron density in the region following the DL. Thus, $n_{\rm w2} = n_{\rm w1} \exp\left[-\frac{\Delta V_{\rm DL}}{T_{\rm w1}}\right]$.

As an application of the above ideas, one may consider the data of ≃ 2 mTorr in Fig. 9(b). In the neighbourhood of $z_1$, P1 has an *average* density $n_1 \simeq 7\times10^{10}$ cm$^{-3}$, $T_{\rm e1} \simeq 25$ eV and $V_{\rm p1} \simeq 93$ V. P2 has $n_2 \simeq 10^{11}$ cm$^{-3}$, $T_{\rm e2} \simeq 13$ eV and $V_{\rm p2} \simeq 49$. V. The difference between $V_{\rm p1}$ and $V_{\rm p2}$ gives, $\Delta V_{\rm DL} \simeq 44$ V. The warm population $P_{\rm we}$ detected after the DL or $z \simeq z_2$ has density $n_{\rm w} \simeq 5.8\times10^{9}$ cm$^{-3}$ and $T_{\rm w} \simeq 47$ eV.

Using $\frac{\Delta V_{\rm DL}}{T_{\rm e1}} \simeq 1.76$, one obtains, $F \simeq 0.318$ and $E_{\rm T} \simeq 23.26$ eV, which yields, $<\epsilon_{\rm PD}> = \frac{E_{\rm T}}{F} \simeq 73$ eV and $n_{\rm w1} \simeq n_{\rm DL} \simeq F\,n_1 \simeq 2.21\times10^{10}$ cm$^{-3}$ (with $n_1 \simeq 7\times0^{10}$ cm$^{-3}$). Finally, $T_{\rm w1} = \frac{2}{3} <\epsilon_{\rm PD}> \simeq 48.7$ eV, which is close to $T_{\rm w}$ (≃ 47 eV), the temperature of the warm population, $P_{\rm we}$ in the experiments. $n_{\rm w2}$, the warm electron density after the DL is given as $n_{\rm w2} = \exp\left[-\frac{\Delta V_{\rm DL}}{T_{\rm w1}}\right] n_{\rm w1} \simeq 0.40\, n_{\rm w1} \simeq 8.4\times10^{9}$ cm$^{-3}$, which matches reasonably with the density, $n_{\rm w}$ ($\simeq 5.8\times10^9$ cm$^{-3}$) of $P_{we}$. To conclude, one may associate the population $P_{we1}$ with $P_{we}$, the population observed in the experiments.

It is seen from above that about $\simeq$ 40 % of the population, $P_{we1}$ would be able to cross the barrier at the DL. The latter was detected as $P_{we}$, in the experiments. It is worth noting that it was possible to detect the population, $P_{we}$ *only* because it constitutes a *high temperature population in the presence of the low temperature electrons* of $P_2$. On the other hand, even though a fraction, $\simeq$ 60 % of $P_{we1}$ would still be present in the region before the DL, it would be difficult to detect such a population in the midst of the warm electrons of $P_1$ (with $T_{e1} \simeq 25$ eV) using LPs at least.

## 4 Comparison of CEPS and GEPS Plasma Characteristics

It was stated in Section 1 that the experiments with the GEPS were conceived and undertaken because it was not possible to probe the plasma close to the source region of the CEPS on account of the extremely harsh plasma conditions there. The relevant results and their analyses have been presented in detail above. In what follows, a comparison of the key features of the CEPS and GEPS plasmas is presented, albeit, only within the EC (4 cm $< z <$ 50 cm).

The primary difference between the CEPS and the GEPS is in their magnetic field configurations. The CEPS uses the MF1 configuration,[31] that provides additional confinement to the plasma electrons in the neighbourhood of the ECR zone allowing electrons to be heated to higher temperature and densities. The GEPS on the other hand, uses the MF2 field topology (see Figs. 2 and 4), which is simpler and offers no additional confinement to the plasma electrons.

Briefly, the results indicate that within 4 cm $< z <$ 50 cm, the plasma parameters for the two setups are approximately comparable over a wide range of pressures, particularly for large $z$. The main difference in the plasma properties for the two setups is evident from the RFEA results, where the energies of emerging ions in the CEPS are significantly higher than those from the GEPS. Plasma parameters and ion energies from both setups are displayed in Figs. (14 – 17) and Table 2.

***Comparison of Plasma Parameters*:** From Figs. 14, one sees as expected that for *both* sources, the plasma density at $z \approx 4$ cm, increases with the pressure up to ≈ 6mTorr. After ≈ 6mTorr however, the density falls for MF1 and saturates for MF2, indicating a shortfall of power with increasing pressure (more so for MF1 than MF2). Nevertheless, the density for MF1 at $z \approx 4$ cm, is consistently higher than for MF2 by a factor of ≈ 2 (or more) across all pressures.

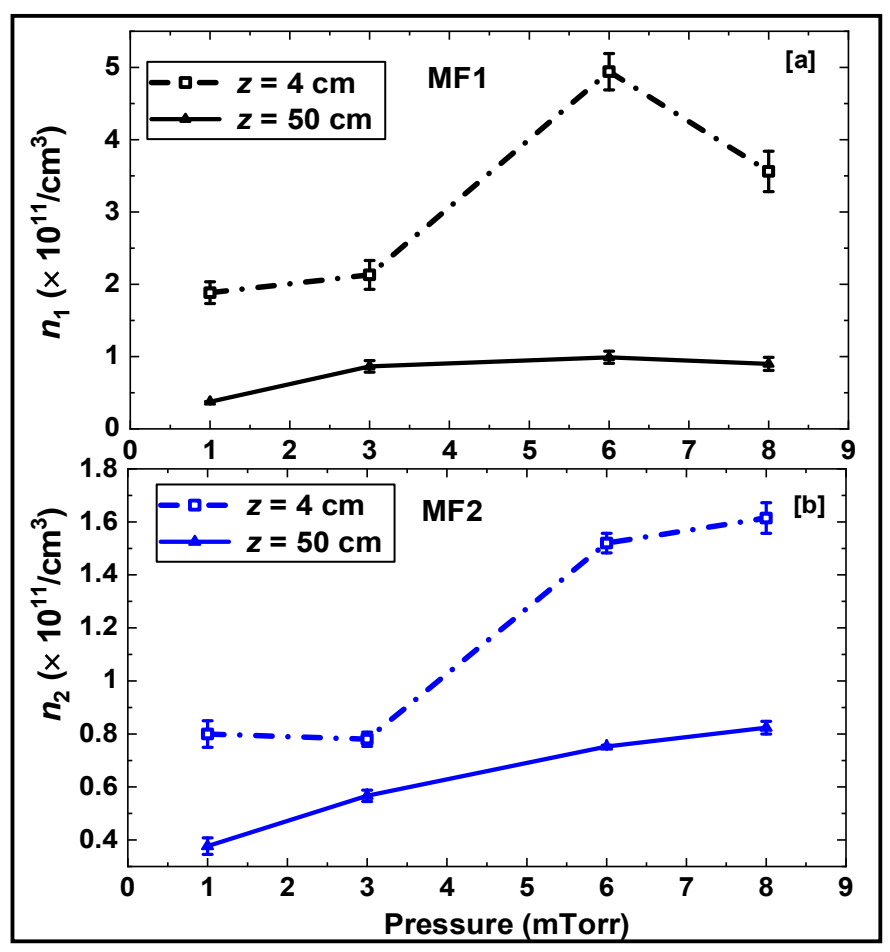

**FIG. 14. Comparison of plasma density with the pressure for both the configurations (a) MF1 (CEPS) and (b) MF2 (GEPS).**

The plasma density at $z \approx 50$ cm on the other hand, is comparable for both sources, tending towards saturation at higher pressures. One gets intermediate values of the density for in between values of $z$. From Figs. 15 (a) and (b) one sees that $T_e$ decreases with the pressure and that the temperature ranges for MF1 and MF2 are comparable. It is noticeable that in the downstream region the temperatures for both magnetic field configurations are fairly low in the range of ≈ 2 – 3 eV (at ≈ 1 mTorr) to about ≈ 0.5 eV (at ≈ 8 mTorr). It is also noteworthy that by adjusting the location and the pressure, one may adjust $T_e$ to any desired value. Figs. 16

(a) and (b) compare the plasma potential $V_p$ for the two field configurations. One notes that at $z \approx 4$ cm, the potential for MF1 is consistently higher by about ≈ 20 V (≈ 5 V) at ≈ 1 mTorr (≈ 8 mTorr). The potential profiles at $z \approx 50$ cm are practically identical over the entire pressure range varying between ≈ 10 – 5 V over the entire range of pressures.

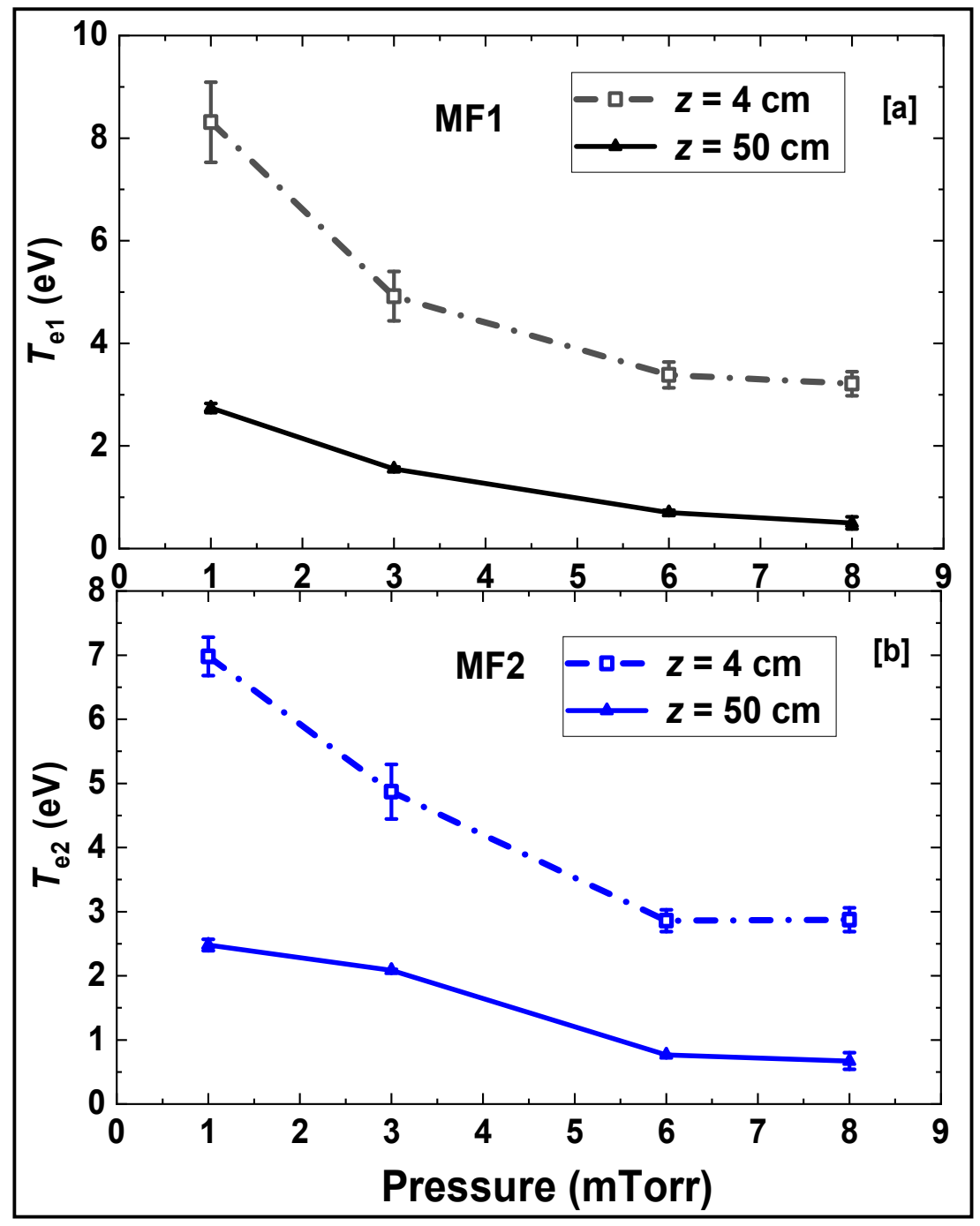

**FIG. 15. Similar figure as in Fig. 14 for the bulk electron temperature**

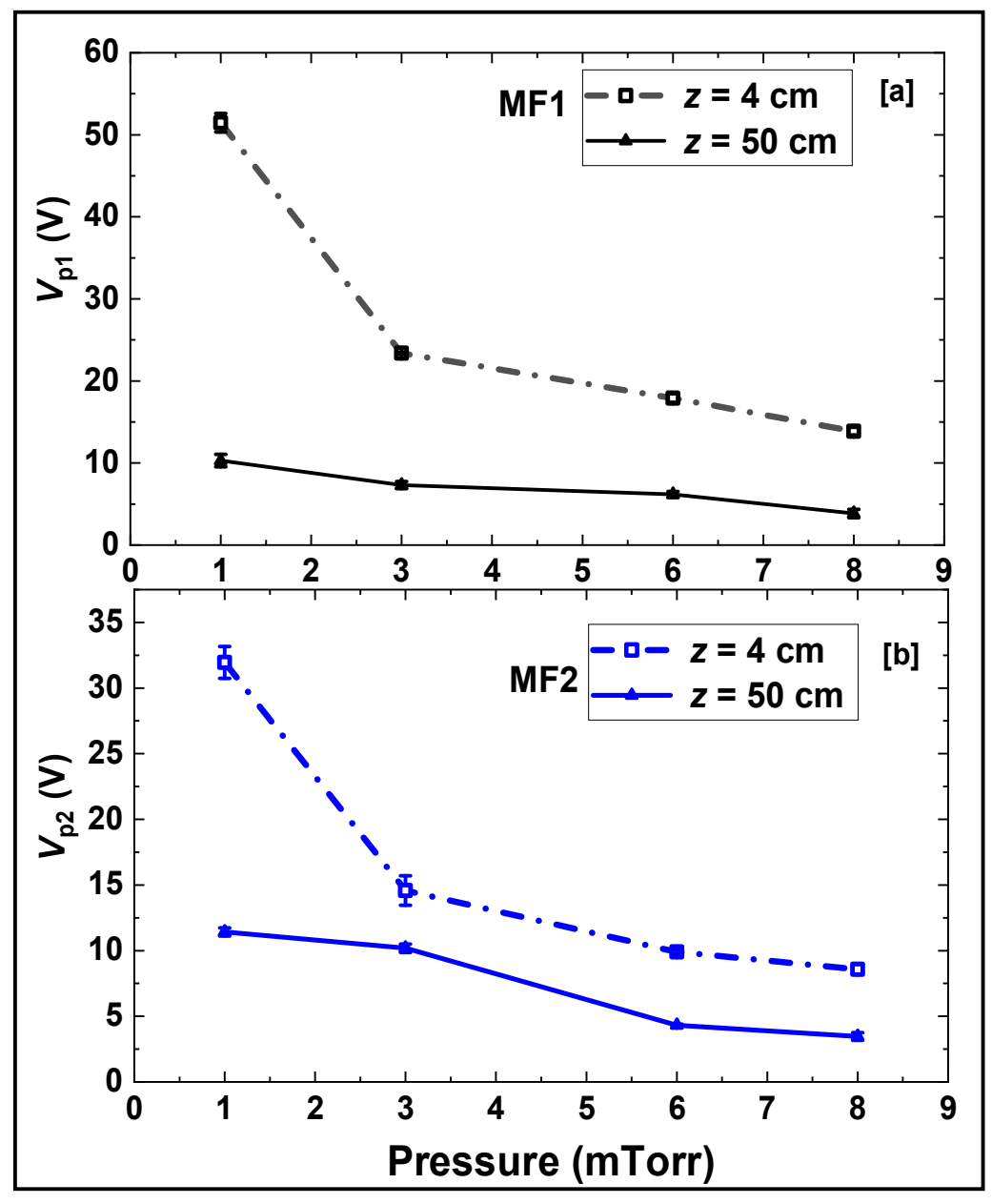

FIG. 16. Similar figure as in Fig. 14 for the plasma potential

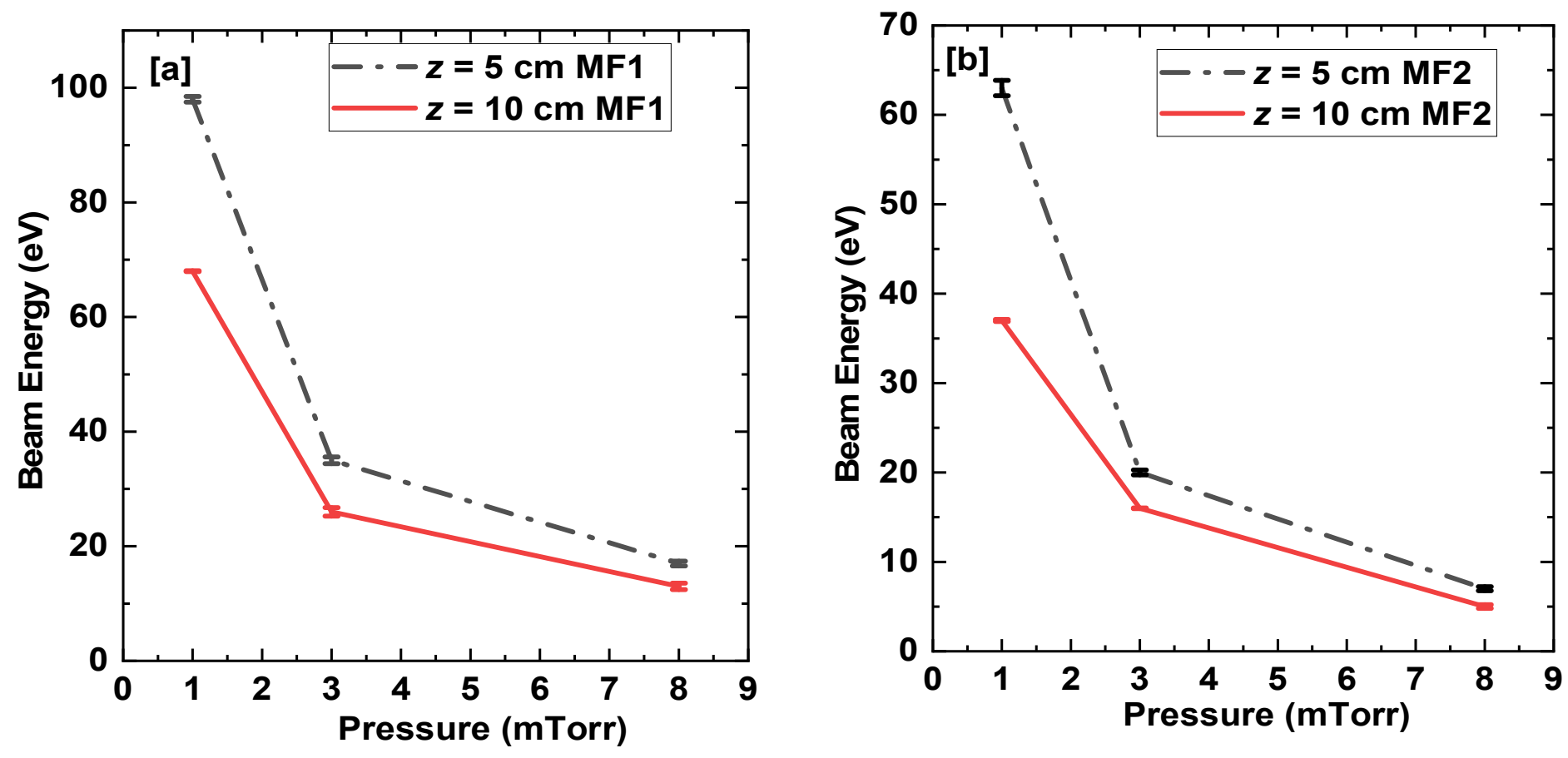

FIG. 17. Comparison of ion energies for the configurations (a) MF1 and (b) MF2 for $z \approx 5$cm and $z \approx 10$ cm.

The ion energies at two locations (see Figs. 17) show the energies for MF1 to be consistently higher than for MF2. A more detailed comparison is given in Table 2 along with the energy spread. The Table reproduces the same features as seen from Fig. 17. In light of the results presented in this work, it is seen that such high ion energies are a consequence of the added confinement by the magnetic field in the ECR zone and DL formation in the PSS of the CEPS. Finally, it is worth noting that the GEPS is actually a *generic source* and that its performance would be comparable to any *typical* ECR source. One may thus conclude that compared to a typical ECR source, the CEPS is superior, both by way of weight and compactness, as well as performance as a plasma source.

**Table 2: Comparison of ion energies (with SD) and energy spreads for MF1 and MF2 at two axial locations**

| | | **$z \approx 5$ cm** | **$z \approx 10$ cm** | **$z \approx 5$ cm** | **$z \approx 10$ cm** |
|---|---|---|---|---|---|
| | | ***Mean Energy*** **[eV]** | ***Mean Energy*** **[eV]** | ***Energy spread*** **[eV]** | ***Energy spread*** **[eV]** |
| **$p \approx 1$ mTorr** | **MF1** | 98±0.5 | 68±0.08 | 12 | 11 |
| | **MF2** | 63±0.85 | 37±0.12 | 22 | 17 |
| **$p \approx 3$ mTorr** | **MF1** | 35±0.58 | 26±0.74 | 16 | 21 |
| | **MF2** | 20±0.27 | 16±0.041 | 25 | 29 |
| **$p \approx 8$ mTorr** | **MF1** | 17±0.42 | 13±0.55 | 9 | 9 |
| | **MF2** | 7±0.23 | 5±0.21 | 12 | 9 |

## 5 Conclusions

This paper presents experiments with a *generic* ECR *plasma source* (GEPS), which offer deeper insight into the properties of plasma produced by the *compact* ECR *plasma source* (CEPS) that was investigated recently.[31] The GEPS produces similar plasma features as the CEPS, albeit less harsh, which allows one to probe the plasma in the source region in some depth. Towards this end, the paper succeeds in providing detailed description of some features like the formation of the second ionization zone close to the ECR zone, initiation of a double layer, production of the warm electron population, etc. Ion with energies in the range ≈ 65 – 7 eV are also detected in front of the source confirming the action of the DL. The paper attempts to analyse and understand these findings in terms of a theoretical model based primarily on the fluid equation for ions. A comparison of the CEPS and GEPS plasmas reveals that the CEPS produces higher density and higher ion energies close to the source exit. However, far away from the source mouth the two sources have similar features determined mainly by the pressure, and the flow properties of the plasma along the magnetic field.

**Acknowledgments**

The authors wish to express gratitude Dr Anshu Verma, Mr A J Josekutty, Mr Hasmukh Kabariya, Ms. Shweta Sharma and Mr. Ashik Basu Mallick for their valuable support in carrying out the experiments and to Prof. P. K. Kaw, Mr. Arun Chakraborty, Prof. Mainak Bandyopadhyay and Prof. Mahendrajit Singh of IPR Gandhinagar for fruitful discussions. One of the authors (SB) is supported by the JRG Program (APCTP) through the S&T Promotion Fund, Lottery Fund (KG) and Korean Local Governments - Gyeongsangbuk-do Province and Pohang City. The authors are also thankful to the Central Research Facility, IIT Delhi for providing liquid nitrogen for the experiments and helping in probe fabrication. This work was supported in part by the Board of Research for Fusion Science and Technology (BFRST) through Memorandum of Understanding (MoU) with the Institute for Plasma Research (IPR) Gandhinagar, Gujarat, India, under Project (IPR ref: IPR/ADMN/10/2015 dated October 2015, IITD ref: RP RP03115G) for the hydrogen experiments undertaken in this manuscript); and in part by the Department of Science and Technology, Government of India for the development of the CEPS.